\documentclass[%
 reprint,prb,
 amsmath,amssymb,
 aps,
floatfix,superscriptaddress
]{revtex4-2}
\usepackage{comment}
\usepackage{mathrsfs}
\usepackage{hyperref}
\usepackage{graphicx}
\usepackage{dcolumn}
\usepackage{bm}
\usepackage{color}
\usepackage{dsfont}
\usepackage{xcolor}
\usepackage[caption=false]{subfig}
\usepackage{appendix}
\usepackage{chngcntr}
\usepackage{natbib}

\hypersetup{
    colorlinks=true,   
    linkcolor=black,   
    citecolor=blue,    
    urlcolor=blue,     
}

\begin{document}

\preprint{APS/123-QED}

\title{
A non-equilibrium quantum transport framework for spintronic devices with dynamical correlations}

\author{Declan Nell}
\affiliation{School of Physics and CRANN Institute, Trinity College Dublin, The University of Dublin, Dublin 2, Ireland}
\author{Milos Radonjic}
\affiliation{Institute of Physics Belgrade, University of Belgrade, 11000 Belgrade, Serbia}
\author{Ivan Rungger}
\affiliation{National Physical Laboratory, Hampton Road, Teddington TW11 0LW, United Kingdom}
\author{Liviu Chioncel}
\affiliation{Theoretical Physics III, Center for Electronic Correlations and Magnetism,
Institute of Physics, and
Augsburg Center for Innovative Technologies, University of Augsburg, 86135 Augsburg, Germany}
\author{Stefano Sanvito}
\affiliation{School of Physics and CRANN Institute, Trinity College Dublin, The University of Dublin, Dublin 2, Ireland}
\author{Andrea Droghetti}
\email[]{andrea.droghetti@unive.it}
\affiliation{School of Physics and CRANN Institute, Trinity College Dublin, The University of Dublin, Dublin 2, Ireland}
\affiliation{Department of Molecular Sciences and Nanosystems, Ca' Foscari University of Venice, via Torino 155, 30170, Mestre-Venice, Italy}

\begin{abstract}
Two-terminal spintronic devices remain challenging to model under realistic operating conditions, where the interplay of complex electronic structures, correlation effects and bias-driven non-equilibrium dynamics may significantly impact charge and spin transport. Existing {\it ab initio} methods either capture bias-dependent transport but neglect dynamical correlations or include correlations but are restricted to equilibrium or linear-response regimes. To overcome these limitations, we present a framework for steady-state quantum transport, combining density functional theory (DFT), the non-equilibrium Greens' function (NEGF) method, and dynamical mean-field theory (DMFT). The framework is then applied to Cu/Co/vacuum/Cu and an Fe/MgO/Fe tunnel junction. In Co, correlations drive a transition from Fermi-liquid to non-Fermi-liquid behavior under finite bias, due to scattering of electrons with electron–hole pairs. This leads to incoherent contributions to the conductance that are observable in scanning tunneling spectroscopy experiments. In contrast, in the Fe/MgO/Fe junction, correlation effects are weaker: Fe remains close to equilibrium even at large biases. Nevertheless, inelastic scattering can still induce partly incoherent transport that modifies the device's response to the external bias. Overall, our framework provides a route to model spintronic devices beyond single-particle descriptions, while also suggesting new interpretations of experiments.
\end{abstract}

\maketitle

\section{\label{sec:level1}Introduction}
Spintronics exploits spin-dependent transport phenomena in materials, with applications in logic 
and data storage \cite{ts.zu.book}. Typical devices are heterostructures formed by stacking thin 
magnetic and non-magnetic layers. Among these, magnetic tunnel junctions 
(MTJs) \cite{bo.cr.01,bu.zh.01,ma.um.01,Yu_Na_2004} exhibiting tunneling magnetoresistance 
(TMR) \cite{ju.75,mi.te.95,mo.ki.95} are already standard components in commercial read heads
for hard-disk drives. However, the computational description of spintronic devices remains generally 
challenging due to their complex electronic structure, their open nature --- allowing carriers to flow 
in and out of the active region --- and the inherently non-equilibrium character of transport.

\begin{figure}[t]
\centering
\includegraphics[width=0.55\textwidth]
{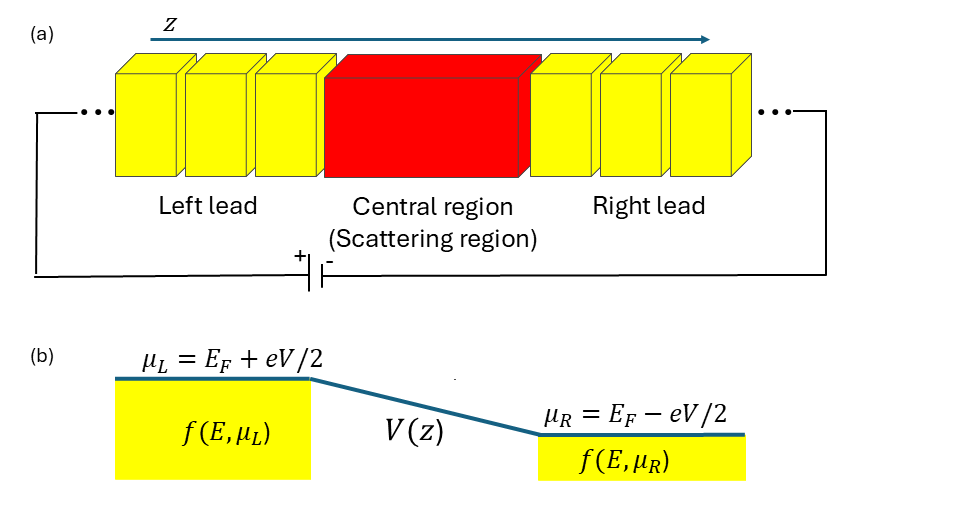} 
\caption{Schematic illustration of a typical two-terminal transport setup. (a) A two-terminal spintronic 
device is modelled as infinite along the transport direction and partitioned into a central (scattering) 
region contacted by semi-infinite metallic leads. (b) Each lead acts as an electronic reservoir. Under 
a finite bias voltage, $V$, the leads individually remain in equilibrium, with their electronic states occupied 
according to the Fermi-Dirac distribution functions, $f_{\mathrm{L(R)}}(E,\mu_\mathrm{L(R)})$, determined
by the relative lead chemical potential $\mathrm{L(R)}$. The electrostatic potential $V(z)$ drops inside 
the central region.}
\label{fig:device} 
\end{figure}

{\it Ab initio} studies typically rely on Kohn-Sham (KS) Density Functional Theory (DFT) \cite{jo.gu.89,kohn.99,jone.15}, 
within either the Local Spin Density Approximation (LSDA) \cite{ba.he.72,vo.wi.80} or the Generalized 
Gradient Approximation (GGA) \cite{pe.ch.92,pe.ch.93,pe.bu.96} for the exchange-correlation functional, 
combined with the Landauer-B\"uttiker (LB) approach \cite{La.57,Bu.86,Bu.88} to quantum transport. In this 
framework, a device is treated as a central open region connected to electrodes that act as charge reservoirs, 
as shown in Fig. \ref{fig:device}(a). The spin-dependent conductance is determined by the transmission of 
spin-polarized KS single-particle states from one electrode, through the central device region, to the opposite 
electrode \cite{sanvito}. Practical implementations include transfer-matrix methods \cite{Wo.Is.02,Wo.Is.02_2}, 
layer Korringa-Kohn-Rostoker approaches \cite{Ma.Zh.99}, and mode-matching techniques \cite{Kh.Br.05}.

The most versatile and widely adopted implementation, however, is based on the non-equilibrium Green's 
function (NEGF) formalism \cite{bookStefanucci, Da.95,book1} in its steady-state formulation. Commonly 
referred to as DFT+NEGF \cite{Ta.Gu.01,Ba.Mo.02,ro.ga.06}, this method enables the calculation of transport 
properties at finite bias, going beyond the linear-response regime assumed in the original LB approach. As a 
result, it provides a device's current-voltage ($I$-$V$) characteristics \cite{Ivan_Stefano_2009, ca.ar.11, sa.ka.12,do.na.14,pa.dl.22} 
as well as other voltage-dependent observables, such as the spin-transfer torque \cite{SL.96,Be.96,My.Ra.99,Ka.Al.00}.

DFT+NEGF relies on the assumption that the KS states provide a reasonable approximation of the excitation 
spectrum of a system. However, beyond issues of formal consistency \cite{Kurth_2017}, this assumption often 
fails even in practice, particularly for the $3d$ ferromagnets (Fe, Ni, Co, and related compounds) widely used 
in spintronic devices. Comparison with photoemission experiments shows that KS-DFT overestimates the spin 
splitting of the $3d$ states and produces majority-spin spectra that are too broad. Moreover, it neglects finite-lifetime 
effects arising from electron-electron interaction, which broaden the electronic bands in both energy and 
momentum \cite{mo.ma.02, Walter_2010}. Finally, the phase-coherent picture of transport in terms of electron 
transmission may break down due to electron correlation effects. For instance, features observed in MTJs' $I$-$V$ 
characteristics have been interpreted as signatures of inelastic electron-electron or electron-magnon scattering
\cite{zh.le.97,ma.yu.05,li.ni.14,ba.dj.09}. While such effects can be included phenomenologically in DFT+NEGF, 
for example via B\"uttiker probes \cite{bu.85,fa.22}, 
a truly {\it ab initio} treatment requires going beyond KS-DFT.

An improved description of the electronic structure of $3d$ metals can be achieved by combining DFT with 
Dynamical Mean-Field Theory (DMFT), in the so-called DFT+DMFT method \cite{me.vo.89,ge.ko.96,li.ka.98,li.ka.01,ko.vo.04,ko.sa.06}. 
Through the incorporation of dynamic, yet local-in-space, correlations, DMFT corrects the overestimation of spin-splitting and 
yields spectral properties in closer agreement with experiment \cite{br.mi.06,gr.ma.07,andrea_sigma_2,ja.dr.23, ne.sa.25}. 
Moreover, DMFT provides a true {\it ab initio} treatment of quasiparticle lifetimes and naturally captures incoherent 
effects arising from electron-electron scattering. 

DMFT can be applied to study transport in devices using a layered formulation \cite{po.no.99,ok.mi.04,free.04, ok.na.05,yu.mo.07,ze.fr.09,va.sa.10, va.sa.12} in the linear-response regime. It can then be combined with 
DFT for {\it ab initio} calculations \cite{andrea_Cu_co,liviu_Cu_Co_dmft,mo.ap.17,ha.ne.24, Ab.Mr.24}, building 
on embedding frameworks originally developed for point contacts \cite{ja.ha.09,ja.ha.10,Ja.15} and molecular junctions \cite{ja.so.13,andrea_ivan_projection,ap.dr.18,ru.ba.19,bh.to.21,gandus2022strongly, ja.18}. The idea is
to apply DMFT to a ``correlated region'', defined within the central device region and hosting significant electron-electron 
interaction. Recently, these DFT+DMFT implementations for transport have also been generalized to ``quasi-equilibrium'' 
conditions \cite{ne.sa.25}, valid when the voltage drop, albeit large in the device, remains small within the 
correlated region.

Genuine non-equilibrium conditions can be tackled by formulating DMFT within the NEGF framework using the 
Keldysh formalism \cite{Ao.Hi.14, Sc.Go.20, Ya.We.23}. Existing implementations, however, are typically designed 
for model bulk systems and focus on transient electron dynamics in the time domain. Studies of steady-state transport 
at finite bias in a device setup remain scarce and are generally limited to single-orbital models, employing 
ans\"atze for the Keldysh component of the many-body self-energy \cite{ok.07,ok.08}, workflows that compute only 
the retarded self-energy \cite{Er.Gu.23}, or master-equation solvers \cite{ar.kn.13}. Implementations for non-equilibrium 
bias-driven transport through realistic materials, described by multiple orbitals per atomic site, are still lacking, leaving electron correlation 
effects in current-carrying spintronic devices largely unexplored.

In this paper, we present a computational framework combining DFT, DMFT and NEGF in the steady-state regime 
to study transport properties of devices beyond the linear-response limit or quasi-equilibrium ans\"atze adopted in 
previous DFT+DMFT transport implementations \cite{ne.sa.25}. Our approach, that we called DFT+NEGF+DMFT, 
enables us to capture correlation effects induced by a finite bias, even in systems where a significant portion of the 
voltage drop occurs within the correlated region, leading to inherently non-equilibrium many-electron behavior. 

Our framework is applied to two prototypical systems. The first one consists of a single Co monolayer 
sandwiched between Cu electrodes, with a gap introduced between one electrode and the Co layer to 
allow for a potential drop under bias. Our results demonstrate that non-equilibrium correlations in Co 
drive a qualitative transition from Fermi-liquid to non-Fermi-liquid behavior, arising from the inelastic 
scattering of electrons with collective electron-hole excitations. This reshapes the DFT-KS electronic 
structure and gives rise to incoherent contributions to the conductance, that can be measured through 
scanning tunneling spectroscopy (STS).

The second system represents the active part of an Fe/MgO/Fe MTJ. Here, the applied bias is found to have a much weaker effect 
on Fe, validating the quasi-equilibrium approximations employed in previous studies \cite{ne.sa.25}. Nevertheless, 
even in this limit, bias-induced inelastic scattering generates incoherent contributions to the current, which 
impact the device transport properties and performance.

The paper is organized as follows. In Sec.~\ref{section: theory}, we introduce the method and describe 
its implementation. Computational details are provided in Sec.~\ref{section: computational details}. Results 
are presented in Sec.~\ref{section: results}, focusing first on the Co/Cu system in Sec.~\ref{sec:CoCu}, 
and then on the Fe/MgO/Fe MTJ in Sec.~\ref{section: Fe/MgO}. Finally, we draw our conclusions in 
Sec.~\ref{section: conclusion}.

\section{Method and implementation} \label{section: theory}

Our framework applies to the standard setup commonly used in quantum transport simulations of 
two-terminal devices under an applied bias voltage, as described in Sec. \ref{sec:system set}. It is 
formulated within the NEGF formalism in the steady-state limit, briefly reviewed in Sec. \ref{sec:GreenFunctions}, 
where the relevant notations are defined. The implementation then proceeds through several key steps, 
as follows. 

At the first level, the method relies on DFT+NEGF calculations to provide an effective single-particle 
description of the device electronic structure, namely a static mean-field treatment under bias. This 
standard methodology is summarized in Sec. \ref{sec.DFTNEGF}. From this starting point, we employ 
an embedding algorithm to extract a correlated subsystem via a projection scheme, detailed in 
Secs. \ref{section: projection} and \ref{sec: embeddingC}. Within this subsystem, electron-electron interactions 
are treated beyond the static mean-field level. Conceptually, our embedding procedure can be viewed as 
a generalization of the band down-folding approaches commonly employed in bulk DFT+DMFT 
(e.g. Ref. \cite{PhysRevB.74.125120}), adapted here to treat devices.

The central development of this work is the introduction of a practical finite-bias DMFT scheme that 
extends the equilibrium approach of Refs. \cite{andrea_sigma_2, andrea_Cu_co, ne.sa.25} to non-equilibrium 
conditions, as explained in Sec. \ref{section: DMFT cycle}. This is achieved by employing a suitable 
non-equilibrium impurity solver (Sec. \ref{section: impurity solver}), which provides the many-body 
self-energies of the correlated subspace under bias. The self-energies are then embedded back into 
the full system (Sec. \ref{sec: back}), enabling the calculations of observables such as the charge 
density, spectral functions, and in particular the steady-state current through the device 
(Sec. \ref{sec.current}). 

The framework is implemented in the {\sc Smeagol} quantum transport package \cite{ro.ga.06}, 
which is interfaced with the {\sc Siesta} DFT code \cite{so.ar.02,Siesta2020}.

\subsection{\label{sec:system set}System setup}

A two-terminal spintronic device, such as a magnetic tunnel junction (MTJ), is modeled as infinite 
along the transport ($z$ by convention) direction and divided into a central scattering region connected 
to semi-infinite metallic leads on the left (L) and right (R) hand side \cite{book1}, as shown in 
Fig.~\ref{fig:device}(a). Periodic boundary conditions are applied in the transverse $x$-$y$ plane, with 
transverse wave vector $\boldsymbol{k} = (k_x, k_y)$ spanning the two-dimensional Brillouin zone (BZ).

The partition into leads and the central region is implemented by using a linear combination of 
atomic orbitals (LCAO) basis set, in which each orbital is assigned to a specific part of the device. 
These basis orbitals are generally non-orthogonal. In principle, changes in the electronic structure 
of the central region feed back into the leads and vice versa, giving rise to interface effects. To 
capture these effects, a sufficient portion of each lead is included within the central region itself. As 
illustrated in Fig.~\ref{fig:projection}(a), the formal lead/central region boundary is therefore not placed 
at the physical interface but rather a few layers deeper inside the leads, where the electronic structure 
has already relaxed to its bulk configuration.

The central region constitutes the active part of the device. In quantum transport calculations, it is 
typically assumed that electronic correlation effects are confined to this region, while the leads are 
treated within an effective single-particle description \cite{haug1996quantum, MW_current, bookStefanucci}. 
Thus, each lead acts as an electronic reservoir, characterized by its chemical potential, $\mu_\mathrm{L/R}$, 
and inverse temperature, $\beta_{\mathrm{L/R}}$. When both leads have identical chemical potentials 
and temperatures --- that is, $\mu_L = \mu_R = E_{\mathrm{F}}$ and $\beta_{\mathrm{L}}=\beta_{\mathrm{R}}$ ---
the system is in thermodynamic equilibrium, described by the grand canonical ensemble, with $E_{\mathrm{F}}$ 
denoting the common Fermi energy.

The effect of a finite bias voltage, $V$, is modeled by shifting the chemical potentials of the leads such 
that $\mu_\mathrm{L} = E_{\mathrm{F}} + \frac{eV}{2}$ and $\mu_\mathrm{R} = E_{\mathrm{F}} - \frac{eV}{2}$, 
where $e$ is the electron charge. Because of the metallic nature of the leads and the requirement of local 
charge neutrality, this shift corresponds to a rigid displacement of their entire band structures. Each lead 
individually remains in equilibrium, described by the Fermi-Dirac distribution functions 
$f_{\mathrm{L(R)}}(E,\mu_\mathrm{L(R)})=\big[1-e^{\beta(E-\mu_\mathrm{L(R)})}\big]^{-1}$, while the central 
region experiences an electrostatic potential drop [see Fig. \ref{fig:device}(b)]. This drives the system out of 
equilibrium and produces a steady-state charge current flowing through the central region from one lead to 
the other.

\subsection{\label{sec:GreenFunctions} NEGF formalism for the steady state transport}

The quantum transport properties of a two-terminal device in the steady state are computed from 
the spin ($\sigma$)-, energy ($E$)- and $\boldsymbol{k}$-dependent NEGF of the central region. 
Its retarded, advanced, and lesser/greater components are given by \cite{th.ru.08}
\begin{equation}
\begin{split}
      G^{\sigma \; r}(E, \boldsymbol{k}) & = [(E+i0^+)S(\boldsymbol{k}) - H^{\sigma}(\boldsymbol{k})   - \Sigma^{\sigma \; r}_\mathrm{L}(E, \boldsymbol{k})\\
       & - \Sigma^{\sigma \; r}_\mathrm{R}(E, \boldsymbol{k}) - \Sigma_{\textnormal{MB}}^{\sigma \; r}(E, \boldsymbol{k})]^{-1}, \label{Eq: mb_retarded_gf}
\end{split}
\end{equation}
\begin{equation}
G^{\sigma \; a}(E, \boldsymbol{k}) = [G^{\sigma \; r}(E, \boldsymbol{k})]^{\dagger},\label{Eq: mb_advanced_gf}
\end{equation}
and
\begin{equation}
\begin{split}
    G^{\sigma \;  \lessgtr} (E, \boldsymbol{k}) & = G^{\sigma \; r} (E, \boldsymbol{k}) \big[ \Sigma^{\sigma \; \lessgtr}_{L}(E, \boldsymbol{k}) + \Sigma^{\sigma \; \lessgtr}_{R}(E, \boldsymbol{k})\\
    &  + \Sigma_{\textnormal{MB}}^{\sigma\; \lessgtr}(E, \boldsymbol{k})\big] G^{\sigma \; a}(E, \boldsymbol{k}),\label{Eq: mb_lesser_gf}
\end{split}
\end{equation}
respectively. 
These satisfy the relation
\begin{equation}\label{eq: Gr-Ga}
G^{\sigma \; r}(E,\boldsymbol{k})-G^{\sigma \; a}(E,\boldsymbol{k})= G^{\sigma \; >}(E,\boldsymbol{k})-G^{\sigma \; <}(E,\boldsymbol{k})\:,
\end{equation} 
so that only two of them are independent. 

In Eq. (\ref{Eq: mb_retarded_gf}), $H^{\sigma}(\boldsymbol{k})$ is the single-particle Hamiltonian 
of the central region, which also includes the electrostatic potential due to the voltage bias 
applied between the leads, and $S(\boldsymbol{k})$ is the orbital overlap.
$\Sigma_{\textnormal{MB}}^{\sigma \; r/a}(E, \boldsymbol{k})$ and $\Sigma_{\textnormal{MB}}^{\sigma \; \lessgtr}(E, \boldsymbol{k})$ 
in Eq. (\ref{Eq: mb_retarded_gf}) and Eq.~(\ref{Eq: mb_lesser_gf}) are respectively the 
retarded/advanced and lesser/greater many-body self-energies, which account for electron-electron 
interaction. Instead, $\Sigma^{\sigma \; r/a}_\mathrm{L}(E, \boldsymbol{k})$ ($\Sigma^{\sigma \; r/a}_\mathrm{R}(E, \boldsymbol{k})$) 
and $\Sigma^{\sigma \; \lessgtr}_\mathrm{L}(E, \boldsymbol{k})$ ($\Sigma^{\sigma \; \lessgtr}_\mathrm{R}(E, \boldsymbol{k})$)
are the retarded/advanced and lesser/greater embedding self-energies of the left-hand side (right-hand side) 
lead, which describe the effect of the leads on the central region. The self-energies satisfy a relation analogous 
to that of Eq. (\ref{eq: Gr-Ga}). The Hamiltonian, the orbital overlap, the Green's functions, and the self-energies 
are represented as square $N\times N$ matrices, where $N$ is the number of basis orbitals within the central 
region. In principle, all of these matrices, except for $S(\boldsymbol{k})$, depend on the bias voltage, $V$. 
However, the explicit $V$-dependence is omitted for the sake of simplicity and will be indicated only when 
necessary.

Eq.~(\ref{Eq: mb_retarded_gf}) is the Dyson equation. The retarded and advanced Green's 
functions, $G^{\sigma \; r}(E, \boldsymbol{k})$ and $G^{\sigma \; a}(E, \boldsymbol{k}) $, describe 
the energy levels in the central region, given by the spin-resolved spectral function, 
\begin{equation}\label{eq:A}
A^\sigma(E, \boldsymbol{k})=i [G^{\sigma \; r}(E,\boldsymbol{k})-G^{\sigma \; a}(E, \boldsymbol{k})]\:.
\end{equation}
The integral of the diagonal matrix elements of $A^\sigma(E, \boldsymbol{k})$ over the transverse 
wave-number, $\boldsymbol{k}$, gives the density of states (DOS). 

Eq.~(\ref{Eq: mb_lesser_gf}) is sometimes called the Keldysh equation \cite{Jauho2006KeldyshNotes}. It describes the 
electron occupation of the central region. In particular, the integral of 
$G^{\sigma \; <}(E, \boldsymbol{k})$ over the energy and the wave-number gives the density 
matrix,
\begin{equation}
 \rho^\sigma=\frac{1}{2\pi i}\frac{1}{\Omega_\mathbf{k}}\int_\mathrm{BZ} d\boldsymbol{k}\int_{-\infty}^{\infty} dE\; G^{\sigma \;<}(E,\boldsymbol{k}),\label{eq.DFT_rho}
\end{equation}
where $\Omega_\mathbf{k}$ is the volume of the transverse BZ.

The retarded embedding self-energy, $\Sigma^{\sigma \; r}_\mathrm{L(R)}(E, \boldsymbol{k})$, can be 
computed exactly using semi-analytical \cite{ru.sa.08} or iterative \cite{Sancho_1984} schemes. Its real part 
gives the energy shifts of the central-region states, while the imaginary part
\begin{equation}\label{eq: Gamma}
\Gamma^{\sigma}_\mathrm{L(R)}(E, \boldsymbol{k}) = i[\Sigma^{\sigma \; r}_\mathrm{L(R)}(E, \boldsymbol{k}) - \Sigma^{\sigma \;r}_\mathrm{L(R)}(E, \boldsymbol{k})^\dagger]\:,
\end{equation}
describes their broadening due to coupling to the leads. The lesser/greater embedding self-energy 
$\Sigma^{\sigma \; \lessgtr}_\mathrm{L(R)}(E, \boldsymbol{k})$ account for the inflow/outflow of charge 
from the L (R) lead to the central region. They can be obtained via the fluctuation-dissipation theorem \cite{bookStefanucci},
\begin{align}
   & \Sigma_{\mathrm{L(R)}}^{\sigma\; <}(E,\boldsymbol{k})=  if_{\mathrm{L(R)}}(E, \mu_{\mathrm{L(R)}})\Gamma^{\sigma}_{\mathrm{L(R)}}(E,\boldsymbol{k})\:, \label{eqn: fluctuation dissapation}\\
   & \Sigma_{\mathrm{L(R)}}^{\sigma\; >}(E,\boldsymbol{k})=  i\big[f_{\mathrm{L(R)}}(E, \mu_{\mathrm{L(R)}})-1\big]\Gamma^{\sigma}_{\mathrm{L(R)}}(E,\boldsymbol{k})\:,\label{eqn: fluctuation dissapation2}
\end{align}
since the leads are charge reservoirs and remain in a local equilibrium. Upon the application of a bias voltage $V$, 
the embedding self-energies are rigidly shifted relative to their zero-bias values
\begin{equation}
\begin{split}
    \Sigma^{\sigma \; r/a \,(\lessgtr)}_\mathrm{L}(E- eV/2,\boldsymbol{k},V)=&\Sigma^{\sigma \; r/a \, (\lessgtr)}_\mathrm{L}(E,\boldsymbol{k},V=0)\:,\\
        \Sigma^{\sigma \; r/a \,(\lessgtr)}_\mathrm{R}(E+eV/2,\boldsymbol{k},V)=&\Sigma^{\sigma \; r/a \, (\lessgtr)}_\mathrm{R}(E,\boldsymbol{k},V=0)\:,
\end{split}\label{eq.SigmaLead_shift}
\end{equation}
where the explicit $V$-dependence has been re-introduced for clarity.

The many-body self-energy satisfies relations analogous to those in Eqs. (\ref{eqn: fluctuation dissapation}) and 
(\ref{eqn: fluctuation dissapation2}) \cite{andrea_ivan_projection, ne.da.10, book1}:
\begin{align}
&\Sigma^{\sigma \; <}_{\textnormal{MB}}(E,\boldsymbol{k})= i F_{\textnormal{MB}}^{\sigma}(E,\boldsymbol{k})\Gamma_{\textnormal{MB}}^{\sigma}(E,\boldsymbol{k}),\label{eq: fd MB1}\\
&\Sigma^{>}_{\textnormal{MB}}(E,\boldsymbol{k})= i[F_{\textnormal{MB}}^{\sigma}(E,\boldsymbol{k})-1]\Gamma_{\textnormal{MB}}(E,\boldsymbol{k}),\label{eq: fd MB2}
\end{align}
where $\Gamma^{\sigma}_\mathrm{MB}(E, \boldsymbol{k}) = i[\Sigma^{\sigma \; r}_\mathrm{MB}(E, \boldsymbol{k}) - \Sigma^{\sigma \;r}_\mathrm{MB}(E, \boldsymbol{k})^\dagger]$ describes the broadening of the spectral function due to 
electron-electron interaction, and $F_{\textnormal{MB}}^{\sigma}(E)$ denotes the occupation matrix of the 
central device region. However, unlike Eqs.~(\ref{eqn: fluctuation dissapation}) and (\ref{eqn: fluctuation dissapation2}), 
$F_{\textnormal{MB}}^{\sigma}(E,\boldsymbol{k})$ generally differs from the Fermi function and reduces to it only at 
zero bias, namely in equilibrium.

Thus, in contrast to the leads, where the fluctuation-dissipation theorem provides a closed form for the retarded 
and lesser/greater self-energies, the many-body self-energies out of equilibrium must be calculated directly,
accounting for the non-equilibrium occupation of the central region. This represents a significant challenge. 
In first-principles calculations, it has been addressed in a few works within the GW or three-body approximations 
of many-body perturbation theory, though these approaches have mostly been applied to molecular 
junctions \cite{th.Ru.07, th.ru.08} or quasi-one-dimensional systems \cite{fe.ca.05_2, fe.ca.05, De.Ca.24}. 
In contrast, DFT+NEGF+DMFT studies of spintronic devices have largely been limited to zero-bias or special 
conditions, where the fluctuation-dissipation theorem is effectively recovered \cite{andrea_sigma_2, andrea_Cu_co}. 
This motivates the need for a general implementation valid under arbitrary non-equilibrium conditions, as presented here.

\subsection{DFT+NEGF}\label{sec.DFTNEGF}
DFT+NEGF neglects the retarded and lesser many-body self-energies and employs the KS 
Hamiltonian, $H^{\sigma}_\mathrm{KS}(\boldsymbol{k})$, as the single-particle Hamiltonian of 
the central region in Eq.~(\ref{Eq: mb_retarded_gf}). The embedding self-energies are likewise 
constructed from the corresponding KS Hamiltonians of the lead materials \cite{rungger_2009}. 
The exchange-correlation potential is usually treated within LSDA \cite{ba.he.72,vo.wi.80} or 
GGA \cite{pe.ch.92,pe.ch.93,pe.bu.96} and acts as a static mean-field interaction. For ferromagnetic 
metals, relevant to the spintronic devices studied in this work, this yields an effective Stoner-like 
description of magnetism \cite{zeller}.
The Green's functions in DFT+NEGF are referred to as KS Green's functions and, throughout 
this paper, are denoted by lowercase symbols $g^{\sigma \; r/a(\lessgtr)}$ to distinguish them 
from the many-body Green's functions introduced in Eqs.~(\ref{Eq: mb_retarded_gf}), 
(\ref{Eq: mb_advanced_gf}), and (\ref{Eq: mb_lesser_gf}).

The KS Hamiltonian of the central region depends functionally on the charge density $\rho^\sigma$, 
defined in Eq. (\ref{eq.DFT_rho}) via the lesser Green's function. Consequently, Eqs. (\ref{Eq: mb_retarded_gf}), 
(\ref{Eq: mb_lesser_gf}), and (\ref{eq.DFT_rho}) must be solved in a self-consistent manner. At finite bias, 
the electrostatic potential across the central region is obtained by solving the Poisson equation for the 
charge density at each iteration step and with the boundary conditions of Eq. (\ref{eq.SigmaLead_shift}) 
\cite{ro.ga.06}. In certain systems, such MTJs, where the voltage drop occurs almost entirely across the 
insulating barrier, the bias dependence can alternatively be incorporated in a non-self-consistent way by 
adding a linear ramp potential to the Hamiltonian. This will be the approach followed in this work, as 
discussed in Sec. \ref{section: computational details}.

\subsection{DMFT combined with DFT+NEGF}\label{section: DMFT}

\begin{figure}[t]
\centering
\includegraphics[width=0.55\textwidth]{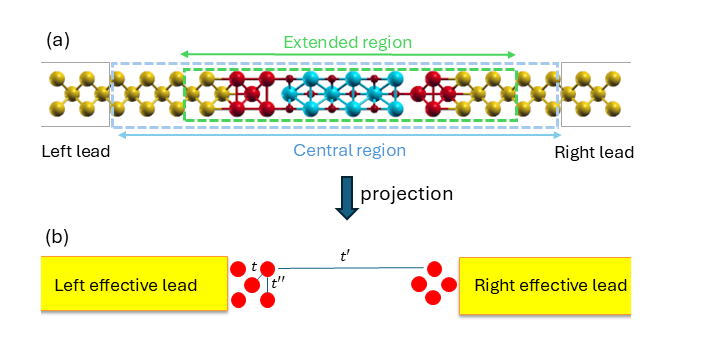} 
\caption{The device setup used in this work. (a) The Fe/MgO/Fe junction sandwiched between simple metallic 
electrodes, introduced in Sec.~\ref{section: Fe/MgO}, shown as a prototypical example of a two-terminal 
device for DFT+NEGF calculations. Fe, Mg, and O atoms are represented by large red, cyan, and small red 
spheres, respectively, while the lead atoms are shown as yellow spheres. The Fe atoms define the correlated 
subspace. The formal lead/central region boundary is placed a few layers deeper inside the leads to ensure 
that interface effects are considered and that the correlated subspace does not interact directly with the leads, 
as required for Eq. (\ref{eq:Sigmaaibath}) to hold. The ``extended region'' is indicated by the green dashed box. 
Atoms outside this region are not coupled to $\mathcal{C}$ due to the finite spatial extent of the basis orbitals. 
(b) The projection maps  the original device onto an effective one, with renormalized hopping between correlated 
atoms and effective coupling to the leads.}
\label{fig:projection} 
\end{figure}

\subsubsection{Projection onto the correlated subspace}\label{section: projection}

The calculation of the many-body self-energy for all the $N$ basis orbitals of a device's central region 
is practically infeasible. Often, this is also unnecessary, since in many cases electron-electron interaction 
beyond the static-mean field KS description is significant only for a few atoms or orbitals within the device. 
Therefore, we carry out a projection of the full central region onto a much smaller subspace, called the 
correlated subspace $\mathcal{C}$, which is spanned by a limited number of orbitals. The orthogonal 
complement of $\mathcal{C}$ within the central region is referred to as the ``bath'', $\mathcal{B}$, and 
remains described at the KS level. Then, $\mathcal{B}$ can be further separated into two new effective 
leads, coupled to the correlated subspace $\mathcal{C}$, thereby providing a mapping from the original 
two-terminal device to a new, effective one, as shown in Fig. \ref{fig:projection}. DMFT will then be applied 
to treat this new system and the obtained self-energies will finally be back-projected to the entire one 
(see Sec. \ref{sec: back}).

Specifically, in the ferromagnetic transition metals used in spintronic devices, the valence states consist 
of the $4s$, $4p$, and $3d$ orbitals. The $4s$ and $4p$ are delocalized and form broad energy bands, 
so that their electronic behavior is well captured by the effective single-particle KS picture. In contrast, the $3d$ 
orbitals are more localized, energetically centered around the Fermi level, and therefore, moderately correlated. 
As such, they are selected to span $\mathcal{C}$. The size of $\mathcal{C}$ is 
$N_\mathcal{C}=2\times 5 \times N_{\textnormal{TM}}$, where $N_{\textnormal{TM}}$ is the number of correlated 
transition metal atoms within our central region. The factor of 2 accounts for spin, and 5 corresponds to the number 
of $3d$ orbitals per atom. Conversely, the size of the bath is $N_\mathcal{B}=N-N_\mathcal{C}$.

The separation between $\mathcal{C}$ and $\mathcal{B}$ is achieved through a basis transformation that 
returns zero orbital overlap between  $\mathcal{C}$ and $\mathcal{B}$ and also orthonormalizes the correlated 
orbitals inside $\mathcal{C}$, as detailed in Ref. \cite{andrea_ivan_projection}. This orthonormalization is convenient, 
as the impurity solvers employed in DMFT are typically formulated and implemented over an orthogonal basis set. 
Denoting the transformation matrix by $W(\boldsymbol{k})$, as defined in Eq. (10) of Ref. \cite{andrea_ivan_projection}, 
the KS Hamiltonian and overlap matrices are transformed accordingly:
\begin{equation}\begin{split}
\bar{H}_\mathrm{KS}^\sigma(\boldsymbol{k})=&\left( \begin{array}{cc}
 \bar{H}^\sigma_\mathcal{C}(\boldsymbol{k})    &\bar{H}^\sigma_{\mathcal{C},\mathcal{B}} (\boldsymbol{k})          \\
 \bar{H}^\sigma_{\mathcal{B},\mathcal{C}}(\boldsymbol{k}) &   \bar{H}^\sigma_\mathcal{B} (\boldsymbol{k})          \\
\end{array} \right)=\\
=& \;W(\boldsymbol{k})^\dagger H^\sigma_{\mathrm{KS}} (\boldsymbol{k}) W(\boldsymbol{k}), \end{split}
\label{eq:Haibath}
\end{equation}
\begin{equation}\begin{split}
\bar{S}(\boldsymbol{k})=& \left( \begin{array}{cc}
 1 &   0          \\
 0 &   \bar{S}_\mathcal{B} (\boldsymbol{k})          \\
\end{array} \right)=\\
=& \;W(\boldsymbol{k})^\dagger S(\boldsymbol{k}) W(\boldsymbol{k}). \end{split}
\label{eq:Saibath}
\end{equation}
Here, $\bar{H}^{\sigma}_{\mathcal{C}}$ ($\bar{H}^{\sigma}_{\mathcal{B}}$) refers to the KS Hamiltonian 
of $\mathcal{C}$ ($\mathcal{B}$), while $\bar{H}^{\sigma}_{\mathcal{CB}} (\boldsymbol{k})$ and 
$\bar{H}^{\sigma}_{\mathcal{BC}} (\boldsymbol{k})$ are those corresponding to the couplings between 
$\mathcal{C}$ and $\mathcal{B}$. In contrast, the transformed overlap matrix, by construction, has the 
$\mathcal{C}$-block equal to the identity matrix, reflecting the orthonormalization of the transformed basis 
within $\mathcal{C}$, and vanishing off-diagonal blocks, because of the zero-overlap condition between 
$\mathcal{C}$ and $\mathcal{B}$.

In second quantization, the KS Hamiltonian of $\mathcal{C}$ reads
\begin{equation}
    \hat{\bar{H}}^{\sigma}_{\mathcal{C}}(\boldsymbol{k}) = \sum_{i  \lambda_i, \; j \lambda_j} [ \bar{H}^{\sigma}_{\mathcal{C}} (\boldsymbol{k})]_{i \lambda_i, j \lambda_j} \hat{d}^{\dagger}_{i \lambda_i \sigma} \hat{d}_{j \lambda_j \sigma} \:,
    \label{eq:HSS_second_q}
\end{equation}
where $\hat{d}^{\dagger}_{i \lambda \sigma}$ ($\hat{d}_{i \lambda \sigma}$) is the fermionic creation (annihilation) 
operator for an electron with spin $\sigma$ in the $\lambda$'th orbital of the $i$'th atomic site. The indices $i$ and 
$j$ run over all atomic sites with correlated orbitals, while $\lambda_i$ and $\lambda_j$ label the correlated orbitals 
on the $i$'th and $j$'th atoms. For example, in a ferromagnetic transition metal with $N_\mathrm{TM}$ atoms  
having correlated $3d$ orbitals, we have $i,j=1, ...,N_\mathrm{TM}$ and $\lambda_i, \lambda_j=1,...,5$.
The diagonal terms in the Hamiltonian matrix, $[ \bar{H}^{\sigma}_{\mathcal{C}} (\boldsymbol{k})]_{i \lambda_i, i \lambda_i}$, 
represent the orbital energies, while the off-diagonal ones, $[ \bar{H}^{\sigma}_{\mathcal{C}} (\boldsymbol{k})]_{i \lambda_i, j \lambda'_i}$ 
with $i\neq j$ ($i=j$ and $\lambda_i\neq\lambda_j$), describe inter (intra)-atomic hopping integrals for 
atoms within $\mathcal{C}$.

The Hamiltonian in Eq.~(\ref{eq:HSS_second_q}) captures the electronic structure of $\mathcal{C}$ within the 
effective single-particle KS picture of DFT+NEGF. To incorporate for electron-electron interaction, we include a 
multi-orbital Hubbard-like interaction term, a choice justified by the predominantly local nature of the screened 
Coulomb interaction in transition metals. The resulting interacting Hamiltonian for $\mathcal{C}$ is
\begin{equation}
\begin{split}
 \hat{\bar{H}}_{\mathcal{C, U}}(\boldsymbol{k}) &= \sum_{\sigma}  \Big[ \hat{\bar{H}}^{\sigma}_{\mathcal{C}}(\boldsymbol{k})   -\hat{\bar{H}}^\sigma_{\mathcal{C},dc} + \\ +\frac{1}{2}\sum_{\substack{i,\lambda_1,\lambda_2,\lambda_3,\\ \lambda_4, \sigma'}} & 
U_{\lambda_1,\lambda_2,\lambda_3,\lambda_4}  \hat{d}_{i \lambda_1 \sigma}^{\dagger} \hat{d}_{i \lambda_2 \sigma'}^{\dagger} \hat{d}_{i \lambda_4 \sigma'} \hat{d}_{i \lambda_3 \sigma} \Big]\:,
 \end{split}\label{eq: Coulomb H}
\end{equation}
where $U_{\lambda_1,\lambda_2,\lambda_3,\lambda_4}$ is the four-index interaction tensor. This can be 
parameterized in terms of the average effective Coulomb repulsion $U$ and exchange $J$ \cite{pavarini}
\begin{equation}
    U = \frac{1}{(2l + 1)^2} \sum_{\lambda_1, \lambda_2} U_{\lambda_1,\lambda_2,\lambda_1,\lambda_2},
\end{equation}
\begin{equation}
    J = \frac{1}{2l(2l + 1)} \sum_{\lambda_1 \neq \lambda_2} U_{\lambda_1,\lambda_2,\lambda_2,\lambda_1}.
\end{equation}
$\hat{\bar{H}}^\sigma_{\mathcal{C},dc}$ is the double-counting correction, which subtracts the static 
correlation already included at the KS mean-field level in $\hat{\bar{H}}^{\sigma}_{\mathcal{C}}(\boldsymbol{k})$. 
While its exact form is unknown, several approximate schemes have been proposed for bulk systems in 
equilibrium \cite{li.ka.01,ko.sa.06,ka.ul.10,ha.ye.10}. However, for open systems out of equilibrium, the proper 
treatment of double-counting remains an open problem, as further discussed in Sec.~\ref{section: impurity solver}.

The same transformation applied to the Hamiltonian is also applied to the lead self-energy. In particular, 
if the system is constructed such that the leads are not directly coupled to $\mathcal{C}$ --- a condition that 
can be ensured by including a sufficient number of lead layers within the central region, as in Fig. \ref{fig:projection}(b) --- 
the lead self-energy acquires the block form
\begin{equation}\begin{split}
\bar{\Sigma}^{\sigma\; r (<)}_{\mathrm{L(R)}}(E,\boldsymbol{k})=& \left( \begin{array}{cc}
 0 &   0          \\
 0 &   \bar{\Sigma}^{\sigma\; r (<)}_{\mathrm{L(R)}, \mathcal{B}} (\boldsymbol{k})          \\
\end{array} \right) =\\
=& \;W(\boldsymbol{k})^\dagger \Sigma^{\sigma\; r (<)}_{\mathrm{L(R)}}(E,\boldsymbol{k}) W(\boldsymbol{k})\:. \end{split}
\label{eq:Sigmaaibath}
\end{equation}
Finally, the sum, $\bar{\Sigma}_{\mathcal{B}} (\boldsymbol{k})   = \bar{\Sigma}_{\mathrm{L}, \mathcal{B}}(\boldsymbol{k}) 
+\bar{\Sigma}_{\mathrm{R}, \mathcal{B}} (\boldsymbol{k})$ describes the total coupling of the bath orbitals to the device's leads.

\subsubsection{Embedding self-energies for the correlated subspace}\label{sec: embeddingC}
The KS Green's function in the transformed basis acquires the same block form as the KS Hamiltonian 
in Eq. (\ref{eq:Haibath}) and it is defined as
\begin{equation}\scalebox{0.86}{$\begin{split}
   & \begin{pmatrix} 
   \bar{g}^{\sigma \; r }_{\mathcal{C}}(E, \boldsymbol{k})  &  \bar{g}^{\sigma \; r }_{\mathcal{CB}}(E, \boldsymbol{k}) \\
    \bar{g}^{\sigma \; r }_{\mathcal{BC}}(E, \boldsymbol{k})  &  \bar{g}^{\sigma \; r }_{\mathcal{B}}(E, \boldsymbol{k})
    \end{pmatrix}=\\ 
 &   \bigg[ E   
    \left( \begin{array}{cc}
 1 &   0          \\
 0 &   \bar{S}_\mathcal{B} (\boldsymbol{k}) 
 \\
\end{array} \right)-\left( \begin{array}{cc}
 \bar{H}^\sigma_\mathcal{C}(\mathbf{k})    &\bar{H}^\sigma_{\mathcal{C},\mathcal{B}} (\boldsymbol{k})          \\
 \bar{H}^\sigma_{\mathcal{B},\mathcal{C}}(\boldsymbol{k}) &   \bar{H}^\sigma_\mathcal{B} (\boldsymbol{k})          \\
\end{array} \right)-\left(\begin{array}{cc}
 0 &   0          \\
 0 &   \bar{\Sigma}_\mathcal{B} (E,\boldsymbol{k}) 
 \\
\end{array}\right)\bigg]^{-1}.
\end{split}
$}
\end{equation}
Performing the matrix inversion yields the KS Green's function of $\mathcal{C}$
\begin{equation}\label{eq:g_c}
    \bar{g}^{\sigma \;r}_\mathcal{C}(E,\boldsymbol{k})=[E-\bar{H}^\sigma_\mathcal{C}(\boldsymbol{k})-
    \bar{H}^\sigma_{\mathcal{C},\mathcal{B}} (\boldsymbol{k})\bar{\mathfrak{g}}^{\sigma\;r}_\mathcal{B}(\boldsymbol{k})\bar{H}^\sigma_{\mathcal{B},\mathcal{C}} (\boldsymbol{k})]^{-1}\:,
\end{equation}
where 
\begin{equation}\label{eq:gbath}
\bar{\mathfrak{g}}^{\sigma\; r}_\mathcal{B}(E,\boldsymbol{k})=[E\bar{S}_\mathcal{B}(\boldsymbol{k})-\bar{H}^\sigma_{\mathcal{B}}(\boldsymbol{k})-\bar{\Sigma}_\mathcal{B} (E,\boldsymbol{k}) ]^{-1}
\end{equation}
is the retarded Green's function of the bath decoupled from $\mathcal{C}$. 

Following Ref. \cite{andrea_ivan_projection}, the bath can be further subdivided into $N_\mathrm{NI}$ 
basis orbitals that are coupled to $\mathcal{C}$, forming the so-called ``extended region'', and the 
$N_\alpha$ ($N_\beta$) orbitals within the central region on the left (right) of $\mathcal{C}$, which are 
not coupled to $\mathcal{C}$ due to the finite spatial extent of the basis orbitals, as shown schematically 
in Fig. \ref{fig:projection}(a).  Thus, the bath and $\mathcal{C}$-bath coupling Hamiltonians can be written in block forms as

%
\begin{eqnarray}
\bar{H}^\sigma_\mathcal{B}(\boldsymbol{k})&=&\left( \begin{array}{ccc}
 H^\sigma_{\mathrm{\alpha\alpha}}(\boldsymbol{k}) & \bar{H}^\sigma_{\mathrm{\alpha,\mathrm{NI}}}(\boldsymbol{k})                    &  H^\sigma_\mathrm{\alpha \beta}(\boldsymbol{k})               \\
 \bar{H}^{\sigma}_{\mathrm{\alpha,\mathrm{NI}}}(\boldsymbol{k})^\dagger & \bar{H}^\sigma_{\mathrm{NI}}(\boldsymbol{k})                           &  \bar{H}^{\sigma}_{\mathrm{\beta,\mathrm{NI}}}(\boldsymbol{k})^\dagger                \\
 H^\sigma_\mathrm{\alpha \beta}(\boldsymbol{k})^\dagger                          & \bar{H}^\sigma_{\mathrm{\beta,\mathrm{NI}}}(\boldsymbol{k})    & H^\sigma_\mathrm{\beta \beta}(\boldsymbol{k})
\end{array} \right), \label{eq:hB}\\
\bar{H}^\sigma_\mathcal{C,B}(\boldsymbol{k})&=&\left( \begin{array}{ccc}
 0 & \bar{H}^\sigma_{\mathcal{C},\mathrm{NI}}(\boldsymbol{k})                 & 0           
\end{array} \right)\:,
\end{eqnarray}
while the bath overlap matrix, $\bar{S}_\mathcal{B}(\boldsymbol{k})$, and the bath Green's 
function, $\bar{\mathfrak{g}}^{\sigma\; r}_\mathcal{B}(E,\boldsymbol{k})$, can be written in the same form as Eq. \eqref{eq:hB}. 
Note that the $H^\sigma_\mathrm{\alpha \beta}$, $H^\sigma_\mathrm{\alpha \alpha}$, and $H^\sigma_\mathrm{\beta \beta}$ 
components are written without bars [cf. Eq. (16) in Ref. \cite{andrea_ivan_projection}], as they remain 
unchanged under the transformation in Eq. (\ref{eq:Haibath}). 

Next, we assume that $H_{\alpha\beta}^{\sigma}(\boldsymbol{k})=0$, which implies that the $N_\alpha$ orbitals to the 
left of $\mathcal{C}$ are not coupled to the $N_\beta$ orbitals to the right, and 
\begin{equation}
\bar{\Sigma}_\mathcal{B} (E,\boldsymbol{k})= \begin{pmatrix}
 {\Sigma}^\sigma_{\mathrm{\alpha\alpha}}(E, \boldsymbol{k}) & 0         &  0              \\
 0 & 0                           &  0     \\
 0 & 0    & {\Sigma}^\sigma_\mathrm{\beta \beta}(\boldsymbol{k}) .     
\end{pmatrix}
\end{equation}
This condition is always enforced by the user in our calculations. Consequently, Eq. (\ref{eq:g_c}) simplifies to
\begin{equation}\label{eq:gKS_C}
    \bar{g}^{\sigma\; r}_\mathcal{C}(E,\boldsymbol{k})=[E-\bar{H}^\sigma_\mathcal{C}(\boldsymbol{k})-
    \bar{\Sigma}^{\sigma\; r}_{\mathcal{C},\mathrm{NI}}(E,\boldsymbol{k})]^{-1}\:,
\end{equation}
where

\begin{equation}\label{eq:sigmaCNI}
\bar{\Sigma}^{\sigma\; r}_{\mathcal{C},\mathrm{NI}}(E,\boldsymbol{k})= \bar{H}^\sigma_{\mathcal{C},\mathrm{NI}}(\boldsymbol{k})\bar{\mathfrak{g}}^{\sigma\; r}_\mathrm{NI} (E,\boldsymbol{k})\bar{H}^\sigma_{\mathrm{NI},\mathcal{C}}(\boldsymbol{k})^\dagger
\end{equation}
is the effective retarded embedding self-energy for $\mathcal{C}$, with the Green's function of the NI part given by
\begin{equation}\label{eq:gNI}\begin{split}
     &\bar{\mathfrak{g}}^{\sigma\; r}_\mathrm{NI}(E,\boldsymbol{k})=\\ &=[E \bar{S}_\mathrm{NI}(\boldsymbol{k})-\bar{H}^\sigma_\mathrm{NI}(\boldsymbol{k})-\bar{\Sigma}^{\sigma\; r}_{\mathrm{L},\mathrm{NI}}(E,\boldsymbol{k}) -\bar{\Sigma}^{\sigma\; r}_{\mathrm{R},\mathrm{NI}}(E,\boldsymbol{k})]^{-1}.
\end{split}
\end{equation}
Here, $\bar{\Sigma}^{\sigma\; r}_{\mathrm{L},\mathrm{NI}}(E,\boldsymbol{k})$ and 
$\bar{\Sigma}^{\sigma\; r}_{\mathrm{R},\mathrm{NI}}(E,\boldsymbol{k})$ are the embedding self-energies for 
NI from the left ($\alpha$) and right ($\beta$) parts of the bath, respectively. These are defined analogously to 
Eqs. (62) and (63) of Ref. \cite{andrea_ivan_projection} as
\begin{equation}
    \begin{split}
        \bar{\Sigma}^{\sigma\; r}_{\mathrm{L},\mathrm{NI}}(E,\boldsymbol{k}) = & \bar{K}^\sigma_{\mathrm{NI},\alpha} (E,\boldsymbol{k})\mathfrak{g}^{\sigma\; r}_{\alpha\alpha}(E,\boldsymbol{k})\bar{K}^\sigma_{\mathrm{NI},\alpha}(E,\boldsymbol{k})^\dagger,\\
\bar{\Sigma}^{\sigma\; r}_{\mathrm{R},\mathrm{NI}}(E,\boldsymbol{k})=&\bar{K}^\sigma_{\mathrm{NI},\beta}(E,\boldsymbol{k})  \mathfrak{g}^{\sigma\; r}_{\beta\beta}(E,\boldsymbol{k}) \bar{K}^\sigma_{\mathrm{NI},\beta}(E,\boldsymbol{k})^\dagger,
    \end{split}
\end{equation}

where 
\begin{equation}
\mathfrak{g}^{\sigma\, r}_{\alpha\alpha (\beta\beta)}(E,\boldsymbol{k})=[K^\sigma_{\alpha\alpha (\beta\beta)}(E,\boldsymbol{k})-\Sigma^\sigma_{\alpha\alpha (\beta\beta)}(E,\boldsymbol{k})]^{-1}
\end{equation}
is the Green's function of the $\alpha$ ($\beta$) block,
$\bar{K}_{\mathrm{NI},\alpha (\beta)}(E,\boldsymbol{k}) =[E \bar{S}_{\mathrm{NI},\alpha (\beta)}(\boldsymbol{k}) - \bar{H}^\sigma_{\mathrm{NI},\alpha (\beta)}(\boldsymbol{k})]$ are the corresponding coupling matrices between NI and 
the left (right) subsystems and $K^\sigma_{\alpha\alpha (\beta\beta)}(E,\boldsymbol{k})=[E \bar{S}_{\alpha \alpha ( \beta \beta)}(\boldsymbol{k}) - \bar{H}^\sigma_{\alpha \alpha ( \beta \beta)}(\boldsymbol{k})]$  corresponds to the left (right) block.

By substituting $\bar{\mathfrak{g}}^{\sigma\; r}_\mathrm{NI}(E,\boldsymbol{k})$ from Eq. (\ref{eq:gNI}) into the 
expression for $\bar{\Sigma}^\sigma_{\mathcal{C},\mathrm{NI}}(E,\boldsymbol{k})$  in Eq. (\ref{eq:sigmaCNI}) 
and by taking the imaginary part, we obtain the broadening matrices for $\mathcal{C}$ coupled to the new 
effective left and right leads 
\begin{equation}
\begin{split}
\bar{\gamma}_\mathrm{L,\mathcal{C}}(E,\boldsymbol{k})=&\bar{H}_\mathrm{\mathcal{C},NI}(\boldsymbol{k}) \bar{\mathfrak{g}}^{\sigma\;r}_\mathrm{NI}(E,\boldsymbol{k}) \bar{\Gamma}_\mathrm{L,NI}(E,\boldsymbol{k})\\ & \times \bar{\mathfrak{g}}_\mathrm{NI}^{\sigma\;r}(E,\boldsymbol{k})^\dagger\bar{H}_\mathrm{\mathcal{C},NI}(\boldsymbol{k})^\dagger,\\
\end{split}
\end{equation}

\begin{equation}
\begin{split}
\bar{\gamma}_\mathrm{R,\mathcal{C}}(E,\boldsymbol{k})=&\bar{H}_\mathrm{\mathcal{C},NI}(\boldsymbol{k}) \bar{\mathfrak{g}}_\mathrm{NI}^{\sigma\;r}(E,\boldsymbol{k})^\dagger \bar{\Gamma}_\mathrm{R,NI}(E,\boldsymbol{k}))\\
& \times\bar{\mathfrak{g}}^{\sigma\;r}_\mathrm{NI}(E,\boldsymbol{k})\bar{H}_\mathrm{\mathcal{C},NI}(\boldsymbol{k})^\dagger,
\end{split}
\end{equation}
as defined in Eqs. (71) and (72) of Ref. \cite{andrea_ivan_projection}, with 
\begin{eqnarray}
&\bar{\Gamma}_\mathrm{L,NI}(E,\boldsymbol{k}))=i\left[\bar\Sigma^{\sigma\;r}_\mathrm{L,NI}(E,\boldsymbol{k}))-\bar\Sigma^{\sigma\;r}_\mathrm{L,NI}(E,\boldsymbol{k})^\dagger)\right],\\
&\bar{\Gamma}_\mathrm{R,NI}(E,\boldsymbol{k}))=i\left[\bar\Sigma^{\sigma\;r}_\mathrm{R,NI}(E,\boldsymbol{k}))-\bar\Sigma^{\sigma\;r}_\mathrm{R,NI}(E,\boldsymbol{k})^\dagger)\right].
\end{eqnarray}

Finally, the lesser KS Green's function of $\mathcal{C}$ is given by
\begin{equation}
\bar{g}^{\sigma\; <}_\mathcal{C}(E,\boldsymbol{k})=\bar{g}^{\sigma\; r}_\mathcal{C}(E,\boldsymbol{k})
[\bar{\Sigma}^{\sigma\; <}_\mathrm{L,\mathcal{C}}(E,\boldsymbol{k})+ \bar{\Sigma}^{\sigma \;<}_\mathrm{R,\mathcal{C}}(E,\boldsymbol{k})] \bar{g}^{\sigma \;r}_\mathcal{C}(E,\boldsymbol{k})^\dagger, 
\end{equation}
where the lesser embedding self-energy for the left-hand side (right-hand side) effective lead is defined, 
analogously to Eq.~(\ref{eqn: fluctuation dissapation}), as
\begin{equation}\label{eqn: fluctuation dissapation_projected}
    \bar{\Sigma}^{\sigma \; <}_\mathrm{L(R),\mathcal{C}}(E, \boldsymbol{k}) =  i f_{\mathrm{L(R)}}(E, \mu_{\mathrm{L(R)}}) \bar{\gamma}_\mathrm{L(R),\mathcal{C}}(E,\boldsymbol{k})\:.
\end{equation}
Here, we use the chemical potential of the original electrodes. As such, while 
Eq.~(\ref{eqn: fluctuation dissapation_projected}) is always valid in equilibrium, under non-equilibrium conditions 
it will hold only if the applied bias does not drop inside the portion of the device that has been incorporated 
into the new effective leads. This is an approximation inherent to our current implementation and should be 
taken into account, as it may limit its applicability. In principle, it can be easily relaxed, but doing so is involved 
within the present structure of the {\sc Smeagol} code and is left for the future work. For the systems studied 
here the approximation, however, is well justified. \\

\subsubsection{Non-equilibrium DMFT loop}\label{section: DMFT cycle}
The effective device is finally treated within DMFT. The retarded and the lesser Green's functions 
of $\mathcal{C}$ are defined analogously to Eq. (\ref{Eq: mb_retarded_gf}) and (\ref{Eq: mb_lesser_gf})
as, 

\begin{widetext}
\begin{equation}
\bar{G}_{\mathcal{C}}^{\sigma\; r}(E,\boldsymbol{k})=\bar{g}^{\sigma\; r}_{\mathcal{C}}(E,\boldsymbol{k})+\bar{g}^{\sigma\; r}_{\mathcal{C}}(E,\boldsymbol{k})\bar{\Sigma}^{\sigma \;  r}_{\mathcal{C}}(E)\bar{G}_{\mathcal{C}}^{\sigma\; r}(E,\boldsymbol{k})\:,\label{eq.dmft_retarded}
\end{equation}
\begin{equation}
\bar{G}^{\sigma\; <}_\mathcal{C}(E,\boldsymbol{k})=\bar{G}^{\sigma\; r}_\mathcal{C}(E,\boldsymbol{k})[\bar{\Sigma}^{\sigma\; <}_\mathrm{L,\mathcal{C}}(E,\boldsymbol{k})+ \bar{\Sigma}^{\sigma \;<}_\mathrm{R,\mathcal{C}}(E,\boldsymbol{k})+\bar{\Sigma}^{\sigma \;  <}_{\mathcal{C}}(E)] \bar{G}^{\sigma \;r}_\mathcal{C}(E,\boldsymbol{k})^\dagger\:.\label{eq.dmft_lesser}
\end{equation}
\end{widetext}
Here, $\bar{g}^{\sigma\; r}_\mathcal{C}(E,\boldsymbol{k})$ is the KS retarded Green's function defined in Eq. (\ref{eq:gKS_C}) 
and $\bar{\Sigma}^{\sigma \; <}_\mathrm{L(R),\mathcal{C}}(E, \boldsymbol{k})$ is the lesser embedding self-energy 
for the left (right) effective lead in Eq. (\ref{eqn: fluctuation dissapation_projected}). $\bar{\Sigma}^{\sigma \;  r}_{\mathcal{C}}(E)$ 
and $\bar{\Sigma}^{\sigma \;  <}_{\mathcal{C}}(E)$ are the many-body retarded and the lesser self-energies, which are 
$\boldsymbol{k}$-independent and site diagonal due to the locality approximation of DMFT,
\begin{equation}\label{eq: block structure of mb self energy}
 \bar{\Sigma}^{\sigma \;  r(<)}_{\mathcal{C}}(E)= 
 \left( \begin{array}{cccc}
 \bar{\Sigma}^{\sigma\;  r(<)}_{1}(E) &   0 & ... & 0          \\
 0 & \bar{\Sigma}^{\sigma\;  r(<)}_{2}(E) & ... & 0           \\
 0 & 0 & ... & \bar{\Sigma}^{\sigma\;  r(<)}_{N_{\textnormal{TM}}}(E)           \\
\end{array} \right).
\end{equation}
Each $\bar{\Sigma}^{\sigma \; r(<)}_i(E)$ block represents the retarded (lesser) DMFT self-energy for atomic site 
$i$ within the correlated subspace $\mathcal{C}$. The index $i$ runs over the $N_\textnormal{TM}$ transition-metal 
atoms forming the ferromagnetic layers in our spintronic devices. Each block is a $5 \times 5$ matrix, corresponding 
to the five $3d$ orbitals of these atoms.  

The self-energies, $\bar{\Sigma}^{\sigma \;  r}_{\mathcal{C}}(E)$ and $\bar{\Sigma}^{\sigma \;  <}_{\mathcal{C}}(E)$, 
are obtained by mapping the correlated subspace onto a set of auxiliary impurity problems, one for each atom. In 
analogy with DMFT for zero-bias transport \cite{andrea_Cu_co,va.sa.12}, this is achieved by enforcing the self-consistency 
condition that the so-called local Green's function of atom $i$ coincides with the corresponding impurity's Green's function,
\begin{equation}
G^{\sigma \; r(<)}_{\textnormal{imp}, i}(E)=\bar{G}^{\sigma \; r(<)}_{\textnormal{loc}, i}(E)\equiv \frac{1}{\Omega_{\boldsymbol{k}}} \int d\boldsymbol{k}\; \bar{G}^{\sigma \; r(<)}_{\mathcal{C}, i}(E, \boldsymbol{k}). \label{LocalGreen}
\end{equation}
Unlike the zero-bias case, where only the retarded component must be matched, under non-equilibrium 
conditions, the self-consistency condition applies to both the retarded and lesser components due to the 
breakdown of the fluctuation-dissipation theorem (Sec.~\ref{sec:GreenFunctions}), except in certain 
special cases (Sec.~\ref{sec:rigid_shift}).

\begin{figure*}[t!]
\centering
\includegraphics[width=0.9\textwidth]{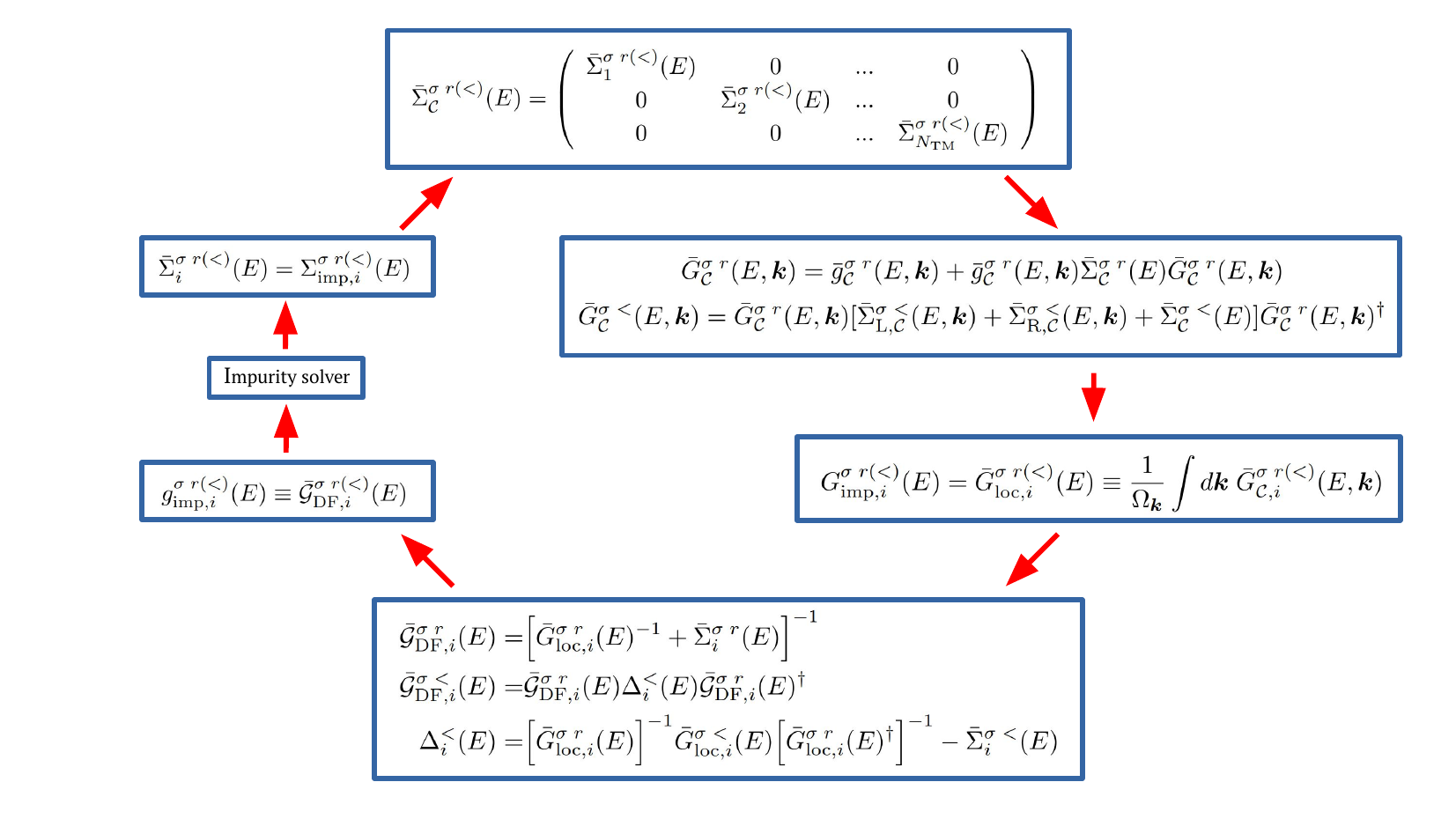} \caption{Schematic representation of the DMFT loop } \label{fig: dmft loop} 
\end{figure*}

In practice, the DMFT self-consistent cycle (schematically illustrated in Fig. \ref{fig: dmft loop}) proceeds through the following main steps:
\begin{itemize} \item[i)] Compute the many-body retarded and the lesser Green's functions of the correlated 
subsystem in Eqs. (\ref{eq.dmft_retarded}) and (\ref{eq.dmft_lesser}). For the first iteration, the many-body 
self-energies $\bar{\Sigma}^{\sigma \;  r}_{\mathcal{C}}(E)$ and $\bar{\Sigma}^{\sigma \;  <}_{\mathcal{C}}(E)$ 
are generally assumed to vanish.

\item[ii)] For each atom $i$, extract the corresponding diagonal block 
$\bar{G}^{\sigma \; r(<)}_{\mathcal{C}, i}(E, \boldsymbol{k})$ from $\bar{G}^{\sigma \; r(<)}_{\mathcal{C}}(E, \boldsymbol{k})$, 
and compute the local retarded (lesser) Green's function.

\item[iii)] Construct the retarded and lesser dynamical mean fields for each atom
\begin{eqnarray} \label{eq: dynamical retarded mean field} 
&\bar{\mathcal{G}}^{\sigma \; r}_{\textnormal{DF}, i}(E) = \Big[\bar{G}^{\sigma \; r}_{\textnormal{loc}, i}(E)^{-1} + \bar{\Sigma}_{i}^{\sigma \; r}(E)\Big]^{-1},
 \\
 &\bar{\mathcal{G}}^{\sigma \; <}_{\textnormal{DF}, i}(E) = \bar{\mathcal{G}}^{\sigma \; r}_{\textnormal{DF}, i}(E) \Delta_{i}^{<}(E) \bar{\mathcal{G}}^{\sigma \; r}_{\textnormal{DF}, i}(E)^{\dagger}, 
\end{eqnarray} 
where
\begin{equation} \label{eq. lesser hybridisation}
\begin{split}
&\Delta_{i}^{<}(E) = \\ &=\Big[\bar{G}^{\sigma \; r}_{\textnormal{loc}, i}(E)\Big]^{-1} \bar{G}^{\sigma \; <}_{\textnormal{loc}, i}(E) \Big[\bar{G}^{\sigma \; r}_{\textnormal{loc}, i}(E)^{\dagger}\Big]^{-1}  - \bar{\Sigma}_{i}^{\sigma \; <}(E).
\end{split}
\end{equation}
is the lesser hybridization function.

\item[iv)] Define the non-interacting impurity's retarded and lesser Green's functions as 
\begin{eqnarray}
& g^{\sigma \; r}_{\textnormal{imp}, i}(E) \equiv \bar{\mathcal{G}}^{\sigma \; r}_{\textnormal{DF}, i}(E)\,,\
& g^{\sigma \; <}_{\textnormal{imp}, i}(E) \equiv \bar{\mathcal{G}}^{\sigma \; <}_{\textnormal{DF}, i}(E).
\label{eq.gimp}
\end{eqnarray}

\item[v)] Solve each impurity problem using an appropriate non-equilibrium impurity solver. The specific solver 
employed in our work is detailed in Sec.~\ref{section: impurity solver}. For each impurity, the solver yields the 
impurity's retarded and lesser many-body self-energies, denoted by $\Sigma^{\sigma \; r}_{\textnormal{imp},i}(E)$ 
and $\Sigma^{\sigma \; <}_{\textnormal{imp},i}(E)$, respectively, and the corresponding impurity's Green's functions,
\begin{align} 
&{G}^{\sigma \; r}_{\mathrm{imp}, i}(E) = \Big[{g}^{\sigma \; r}_{\mathrm{imp}, i}(E)^{-1}  -\Sigma^{\sigma \; r}_{\textnormal{imp},i}(E)\Big]^{-1},
 \\
 &G^{\sigma \; <}_{\mathrm{imp}, i}(E) = G^{\sigma \; r}_{\mathrm{imp}, i}(E) \Big[\Delta_{i}^{<}(E)+\Sigma^{\sigma \; <}_{\textnormal{imp},i}(E)\Big] G^{\sigma \; r}_{\mathrm{imp}, i}(E)^{\dagger}.
\end{align} 

\item[vi)] Update the DMFT retarded and lesser self-energies by setting $\bar{\Sigma}_{i}^{\sigma \; r}(E) = \Sigma^{\sigma \; r}_{\textnormal{imp}, i}(E)$ and $\bar{\Sigma}_{i}^{\sigma \; <}(E) = \Sigma^{\sigma \; <}_{\textnormal{imp}, i}(E)$ in 
Eq. (\ref{eq: block structure of mb self energy}). Return to step (i) and iterate until the self-consistency condition in 
Eq.~(\ref{LocalGreen}) is satisfied.

\end{itemize}

\subsubsection{Rigid shift approximation}\label{sec:rigid_shift}
In certain situations, the full non-equilibrium DMFT procedure described above can be simplified. In particular, 
when the correlated atoms are strongly coupled to one lead and are separated from the opposite lead by a sufficiently 
thick barrier --- so that their interaction with that well-separated lead is negligible --- they can be considered to be in 
equilibrium with the lead they are coupled to. In this limit, the fluctuation-dissipation theorem is effectively recovered.

This allows us to apply the ``rigid shift approximation'', introduced in Ref. \cite{ne.sa.25}. Each block of the 
retarded and lesser correlated subspace's self-energies in Eq.~(\ref{eq: block structure of mb self energy}) 
can be obtained through relations similar to Eq.~(\ref{eq.SigmaLead_shift}) and Eq.~(\ref{eqn: fluctuation dissapation}), 
namely 
\begin{eqnarray}
&\bar{\Sigma}^{\sigma \; r}_{i}(E\pm eV/2,V)=\bar{\Sigma}^{\sigma \; r}_{i}(E,V=0)\:,\label{eq.SigmaDMFT_shift}\\ 
&\bar{\Sigma}^{\sigma \; <}_{i}(E,V)=if_\mathrm{L(R)}(E,\mu_\mathrm{L(R)})\bar{\Gamma}^{\sigma}_{i}(E,V)\:,\label{eq.sigmaDMFT_lesser}
\end{eqnarray}
where $
\bar{\Gamma}^{\sigma}_{i}(E, V) = i[\tilde{\Sigma}^{\sigma \; r}_{i}(E, V) - \bar{\Sigma}^{\sigma \;r}_{i}(E, V)^\dagger]$. 
The minus (plus) sign in Eq.~(\ref{eq.SigmaDMFT_shift}) and the Fermi function
$f_\mathrm{L(R)}$ in Eq.~(\ref{eq.sigmaDMFT_lesser}) are used when the atom $i$ is coupled to the left-hand side 
(right-hand side) lead. By using this approach, only the retarded self-energy $\bar{\Sigma}^{\sigma \; r}_{i}(E,V=0)$ at 
zero-bias needs to be computed; the lesser and retarded components at finite bias can then be obtained directly, 
thus greatly reducing the computational effort. 

In practice, however, we find that a straightforward application of the rigid-shift approximation often leads 
to poor current conservation, as it neglects the change in hybridization between the leads and the correlated 
orbitals induced by the electrostatic potential drop. To address this shortfall, we compute $\tilde{\Sigma}^{\sigma \; r}_{i}(E,V)$ 
using an equilibrium calculation with this electrostatic potential added to the zero-bias Kohn-Sham Hamiltonian 
of the central region, and then obtain the corresponding lesser DMFT self-energy via Eq.~(\ref{eq.sigmaDMFT_lesser}). 

\subsubsection{Back-projection to the CR}\label{sec: back}
Once the DMFT self-energy of $\mathcal{C}$ has been calculated, we can straightforwardly construct 
the self-energy of the full central region as
\begin{equation}
     \bar{\Sigma}_{\textnormal{MB}}^{\sigma \; r(<)}(E)  =
    \begin{pmatrix} 
          \bar{\Sigma}_{\textnormal{MB}, \mathcal{C}}^{\sigma \; r(<)}(E) &       0 \\     0  &     0
    \end{pmatrix}\:, \\
\end{equation}
and apply the inverse transformation of Eq. (\ref{eq:Haibath}) to express the many-body self-energy in the original basis,
\begin{eqnarray}
\Sigma^{\sigma\; r (<)}_\mathrm{MB}(E, \boldsymbol{k})=W(\boldsymbol{k}) ^{-1\,\dagger}\,\bar{\Sigma}^{\sigma\; r (<)}_\mathrm{MB}(E) \, W(\boldsymbol{k})^{-1}.
\label{eq:sigmambinv}
\end{eqnarray}
Then, $\Sigma^{\sigma\; r (<)}_\mathrm{MB}(E, \boldsymbol{k})$ can then be used  to compute the Green's 
functions defined in Eqs. (\ref{Eq: mb_retarded_gf}) and (\ref{Eq: mb_lesser_gf}), and thereby the spectral 
function [Eq. (\ref{eq:A})], the density matrix [Eq. (\ref{eq.DFT_rho})], and the charge current defined in 
Sec. \ref{sec.current}.

\subsubsection{Impurity solver} \label{section: impurity solver}

At equilibrium, the impurity problem is often solved with high accuracy --- up to stochastic noise --- using 
continuous-time quantum Monte Carlo (CTQMC) methods \cite{gu.mi.11}, formulated on the imaginary 
frequency axis. To obtain real-frequency observables such as spectral functions, these results must be 
analytically continued to the real axis, a numerically ill-posed procedure \cite{JARRELL1996133}. Extending CTQMC methods to 
non-equilibrium systems is significantly more challenging, as it requires real-time simulations along the 
Keldysh contour \cite{mu.ra.08,sc.we.08,sc.fa.09,we.ok.09,we.ok.09_2}, a task that suffers from severe 
sign problems and an exponential growth in computational cost with simulation time. Although recent 
advances, such as the steady-state inchworm CTQMC algorithm \cite{Er.Gu.23}, have reduced the computational 
cost by several orders of magnitude and enabled the calculation of frequency-dependent steady-state 
quantities, applications have so far been limited to the single-orbital Anderson impurity model. Unfortunately, 
they remain computationally prohibitive for realistic multi-orbital impurity problems. 

Given our focus on transition metals, which present moderate correlations, we adopt a more computationally 
efficient impurity solver based on second-order perturbation theory in the electron-electron 
interaction \cite{bookStefanucci, st.21, andrea_sigma_2}. This approach can be formulated directly on the 
real energy axis and has been shown in previous studies to produce results in good agreement with experimental 
photoemission spectrum of transition metals thin films \cite{andrea_FeO, andrea_sigma_2}. 

In our calculations, we assume that both the retarded and lesser components of the impurity Green's 
function, as well as the self-energy, are diagonal in orbital space. This significantly reduces the computational 
cost. For the systems considered here, this assumption is well justified due to their cubic symmetry and is routinely utilized in the literature \cite{li.ka.01, he.ne.01}. In less symmetric 
systems, however, the diagonal approximation may not hold, and neglecting orbital off-diagonal elements introduces 
an approximation whose significance must be assessed on a case-by-case basis.

The retarded component of the impurity self-energy contains both first- and second-order terms in the 
Coulomb interaction,
\begin{equation}
    \Sigma^{\sigma \; r }_{\textnormal{imp},i\lambda}(E) \approx \Sigma^{\sigma \; r (1)}_{\textnormal{imp},i\lambda} + \Sigma^{\sigma \; r (2)}_{\textnormal{imp},i\lambda}(E),
\end{equation}
while the lesser component only contains second-order terms,
\begin{equation}
    \Sigma^{\sigma \; < }_{\textnormal{imp},i\lambda}(E) \approx \Sigma^{\sigma \; < (2)}_{\textnormal{imp},i\lambda}(E).
\end{equation}
The subscript $\lambda$ denotes the orbital of the impurity $i$, as in Eq. (\ref{eq: Coulomb H}).

The first order contribution to the retarded self-energy takes the well-known Hartree-Fock form:
\begin{equation}
\Sigma^{\sigma  \; r (1)}_{\textnormal{imp},i\lambda}(E) = \sum_{i\lambda_1 \sigma_1} U_{i\lambda i\lambda_1 i\lambda i\lambda_1}^{\sigma\sigma_1\sigma\sigma_1} n_{i \lambda_1}^{\sigma_1} - \sum_{i\lambda_1} U_{i\lambda i\lambda_1 i\lambda_1 i\lambda}^{\sigma\sigma\sigma\sigma} n_{i \lambda_1}^{\sigma},
\end{equation}
which is local in time and therefore energy-independent. Here, 
$n_{i \lambda}^{\sigma} = -\mathrm{i} \int \frac{dE}{2\pi} g^{\sigma\,<}_\mathrm{imp}{i \lambda}(E)$ denotes the occupation 
of orbital $\lambda$ with spin $\sigma$ on impurity $i$, where $g^{\sigma\,<}_{\mathrm{imp}, i \lambda}(E)$ is the 
corresponding impurity Green's function defined in Eq.~(\ref{eq.gimp}). 

The second-order contributions to the retarded and lesser self-energies are given by 

\begin{widetext}
\begin{eqnarray}
\label{eq:sig2r}
\mathrm{Im}\left[\Sigma^{\sigma \; r (2)}_{\textnormal{imp},i\lambda}(E)\right] &= 
-\pi \sum\limits_{i\lambda_1 i\lambda_2 i\lambda_3\sigma_1}& U_{i\lambda i\lambda_1 i\lambda_2 i\lambda_3}^{\sigma\sigma_1\sigma\sigma_1}U_{i\lambda_3 i\lambda_2 i\lambda_1 i\lambda}^{\sigma_1\sigma\sigma_1\sigma} 
\int \mathrm{d}\varepsilon_1 \mathrm{d} \varepsilon_2 
\rho_{i\lambda_1\sigma_1}(\varepsilon_1) 
\rho_{i\lambda_2\sigma}(\varepsilon_2) 
\rho_{i\lambda_3\sigma_1}(\varepsilon_1+\varepsilon_2-E) \times \\ \nonumber 
&&\{ \tilde{f}_{i\lambda_1\sigma_1}(\varepsilon_1) \tilde{f}_{i\lambda_2\sigma}(\varepsilon_2) + 
\left[ 1-\tilde{f}_{i\lambda_1\sigma_1}(\varepsilon_1) - \tilde{f}_{i\lambda_2\sigma}(\varepsilon_2) \right] \tilde{f}_{i\lambda_3\sigma_1}(\varepsilon_1+\varepsilon_2-E) \}\\ \nonumber
&+\pi \sum\limits_{i\lambda_1 i\lambda_2 i\lambda_3}& U_{i\lambda i\lambda_1 i\lambda_2 i\lambda_3}^{\sigma\sigma\sigma\sigma}U_{i\lambda_2 i\lambda_3 i\lambda_1 i\lambda}^{\sigma\sigma\sigma\sigma} 
\int \mathrm{d}\varepsilon_1 \mathrm{d} \varepsilon_2 
\rho_{i\lambda_1\sigma}(\varepsilon_1+\varepsilon_2-E) 
\rho_{i\lambda_2\sigma}(\varepsilon_2) 
\rho_{i\lambda_3\sigma}(\varepsilon_1) \times \\ \nonumber 
&&\{ \tilde{f}_{i\lambda_2\sigma}(\varepsilon_2) \tilde{f}_{i\lambda_3\sigma}(\varepsilon_1) + 
\left[ 1-\tilde{f}_{i\lambda_2\sigma}(\varepsilon_2) - \tilde{f}_{i\lambda_3\sigma}(\varepsilon_1) \right] \tilde{f}_{i\lambda_1\sigma}(\varepsilon_1+\varepsilon_2-E) \}\\
\label{eq:sig2l}
\Sigma^{\sigma \; < (2)}_{\textnormal{imp},i\lambda}(E) &=
2\pi \mathrm{i} \sum\limits_{i\lambda_1 i\lambda_2 i\lambda_3\sigma_1}& U_{i\lambda i\lambda_1 i\lambda_2 i\lambda_3}^{\sigma\sigma_1\sigma\sigma_1}U_{i\lambda_3 i\lambda_2 i\lambda_1 i\lambda}^{\sigma_1\sigma\sigma_1\sigma} 
\int \mathrm{d}\varepsilon_1 \mathrm{d} \varepsilon_2 
\rho_{i\lambda_1\sigma_1}(\varepsilon_1) 
\rho_{i\lambda_2\sigma}(\varepsilon_2) 
\rho_{i\lambda_3\sigma_1}(\varepsilon_1+\varepsilon_2-E) \times \\ \nonumber 
&&\tilde{f}_{i\lambda_1\sigma_1}(\varepsilon_1) \tilde{f}_{i\lambda_2\sigma}(\varepsilon_2) \left[ 1-\tilde{f}_{i\lambda_3\sigma_1}(\varepsilon_1+\varepsilon_2-E) \right]\\ \nonumber
&-2\pi \mathrm{i} \sum\limits_{i\lambda_1 i\lambda_2 i\lambda_3}& U_{i\lambda i\lambda_1 i\lambda_2 i\lambda_3}^{\sigma\sigma\sigma\sigma}U_{i\lambda_2 i\lambda_3 i\lambda_1 i\lambda}^{\sigma\sigma\sigma\sigma} 
\int \mathrm{d}\varepsilon_1 \mathrm{d} \varepsilon_2 
\rho_{i\lambda_1\sigma}(\varepsilon_1+\varepsilon_2-E) 
\rho_{i\lambda_2\sigma}(\varepsilon_2) 
\rho_{i\lambda_3\sigma}(\varepsilon_1) \times \\ \nonumber 
&&\tilde{f}_{i\lambda_2\sigma}(\varepsilon_2) \tilde{f}_{i\lambda_3\sigma}(\varepsilon_1) \left[ 1-\tilde{f}_{i\lambda_1\sigma}(\varepsilon_1+\varepsilon_2-E) \right],
\end{eqnarray}
\end{widetext}
while the real part of the retarded self-energy is obtained from the Kramers-Kronig relations. 
Here, $\rho_{i\lambda\sigma}(E)=-\frac{1}{\pi} \mathrm{Im} g^{\sigma\,r}_{\mathrm{imp}, i \lambda}(E)$ denotes 
the spectral function of orbital $\lambda$ with spin $\sigma$ on impurity $i$, and \(\tilde{f}_{i\lambda \sigma}(E)\) 
is the corresponding generalized Fermi function defined as
\begin{equation}
\label{eq:ftilde_def}
\tilde{f}_{i\lambda \sigma}(E) = \frac{g^{\sigma<}_{\mathrm{imp}, i \lambda}(E) }{2\pi \rho_{i\lambda\sigma}(E)}.
\end{equation}
The numerical integrations (convolutions) are performed using the uniform mesh and the trapezoid rule integration.

The first-order term corresponds to an on-site energy shift that adds to the double-counting correction in 
Eq.~(\ref{eq: Coulomb H}). In previous implementations~\cite{andrea_sigma_2}, limited to zero bias, the 
sum of these terms was approximated by the effective $U$ potential of the DFT+$U$ formulation by 
Dudarev {\it et al.}~\cite{dudarev}. In the present work, since the proper treatment of double-counting 
corrections in transport calculations at finite bias remains an open problem and the first-order term is 
numerically difficult to converge, we assume that these contributions cancel exactly and therefore 
we neglect both.

The $U$ potential and the second-order self-energy generally have opposite effects in transition-metal 
ferromagnets: the former enhances the spin-splitting already overestimated by DFT, while the latter tends 
to reduce it. Accordingly, in this work the omission of the $U$ potential is compensated by adopting smaller 
effective $U$ and $J$ values in the evaluation of $\Sigma^{\sigma \; r (2)}_{\textnormal{imp},i\lambda}(E)$. 
At zero bias, this approximation yields results consistent with Refs.~\cite{andrea_Cu_co,ne.sa.25} for Fe/MgO 
and Co/Cu junctions, where the first-order term was explicitly included, indicating that both approaches are 
equivalent for our practical purposes.

\subsection{Charge current}\label{sec.current}

The expression for the charge current flowing from the non-interacting left-hand side (right-hand side) 
lead into the CR of a two-terminal device in steady-state conditions was derived by Meir and Wingreen 
in their seminal paper~\cite{MW_current}, and is given by
\begin{widetext}
\begin{equation}
    I_\mathrm{L(R)} = 
    \frac{ie}{h}\frac{1}{\Omega_{\boldsymbol{k}}}
   \sum_\sigma \int  d\boldsymbol{k} \int_{-\infty}^{\infty} 
    dE\,\textnormal{Tr}\Big[f_\mathrm{L(R)}(E, \mu_\mathrm{L(R)})\Gamma^\sigma_\mathrm{L(R)}(E,\boldsymbol{k}) 
    A^\sigma(E,\boldsymbol{k}) 
   + i \Gamma^\sigma_\mathrm{L(R)}(E,\boldsymbol{k})G^{\sigma\,<}(E,\boldsymbol{k})\Big],\label{eq.IL_MW}
\end{equation}
\end{widetext}

where $G^{\sigma\,<}(E,\boldsymbol{k})$, $A^\sigma(E,\boldsymbol{k})$, and 
$\Gamma^\sigma_\mathrm{L(R)}(E,\boldsymbol{k})$ are defined in Eqs. (\ref{Eq: mb_lesser_gf}), (\ref{eq:A}), and 
(\ref{eq: Gamma}), respectively. Although Eq. (\ref{eq.IL_MW}) was originally derived assuming an orthonormal basis, 
it remains valid for the non-orthogonal case as well 
\cite{Th.06}\footnote{Note that in Eq. (45) of Ref. \cite{Th.06}, the Meir-Wingreen formula is recovered for 
a non-orthogonal basis in terms of what the author calls the ``overlap Green's functions'', denoted using 
Gothic symbols. The definition of these Green's functions in Eq. (16) of Ref. \cite{Th.06} coincides with the 
one we use in Eq. (\ref{Eq: mb_retarded_gf}) so that that the application of the Meir-Wingreen formula in 
our work is rigorous.}.

Since charge conservation implies that $I_\mathrm{L} = - I_\mathrm{R}$, the current through the CR 
is usually expressed in the symmetrized form, $I=(I_\mathrm{L}-I_\mathrm{R})/2$, a convention adopted here.
In practice, however, some care is required in the calculations. Although DMFT is a conserving many-body 
theory in the Baym-Kadanoff sense~\cite{ko.sa.06}, ensuring charge conservation in practice can be challenging 
due to the employed approximations (e.g., in the implementation of the impurity solvers) and numerical error 
propagation. Thus, we explicitly verify current conservation in all our calculations, as detailed in 
Sec.~\ref{section: current conservation}, finding generally satisfactory results.

The current can then be expressed as the sum of two contributions, which we refer to as the coherent and incoherent 
parts,
\begin{equation} \label{eq: coherent + incoherent}
I^{\sigma} = I_{\textnormal{C}}^{\sigma} + I_{\textnormal{IC}}^{\sigma}.
\end{equation}
This decomposition follows from Eq. (\ref{eq: Gr-Ga}), together with the expressions for the lesser and 
greater components of the lead and many-body self-energies given in Eqs. (\ref{eqn: fluctuation dissapation}) 
and (\ref{eq: fd MB1}).
The coherent contribution reads
 \begin{equation} 
\begin{split}
    I_{\textnormal{C}}^{\sigma} = \frac{e}{h} \int_{-\infty}^{\infty}\! dE \;
 T^\sigma(E) \big[f_{\textnormal{L}}(E)-f_{\textnormal{R}}(E)\big],\label{eq: coherent current}
\end{split}
\end{equation}
where $T^{\sigma}(E)$ denotes the  transmission coefficient, given by
\begin{equation}\begin{split}
T^\sigma(E)= \frac{1}{\Omega_{\boldsymbol{k}}}\int_\mathrm{BZ} d\boldsymbol{k}\;\textnormal{Tr} & \Big[\Gamma^{\sigma}_\mathrm{L}(E, \boldsymbol{k})  G^{\sigma\; r}(E, \boldsymbol{k})\times\\
 &\Gamma^{\sigma}_\mathrm{R}(E, \boldsymbol{k})G^{\sigma\; r}(E, \boldsymbol{k})^{\dagger}\Big]\:.
\end{split}\label{eq: TRC many body k-dependent}
\end{equation} 
Equation (\ref{eq: coherent current}) is the generalization of the Laudauer-B\"uttiker formula to finite bias (see, for example, 
reference \cite{Da.95,sanvito, ro.ga.06}).
The incoherent contribution, instead, is
\begin{equation} \label{eq: incoherent current k dependent}
    \begin{split}
        I_{\textnormal{IC}}^{\sigma} = &\frac{e}{2h\,\Omega_{\boldsymbol{k}}}\int_\mathrm{BZ} d\boldsymbol{k}\int dE\; \textnormal{Tr}\Big\{\big[f_{\textnormal{L}}(E)-F_{\textnormal{MB}}^{\sigma}(E,\boldsymbol{k})\big]\Gamma_{\textnormal{MB}}^{\sigma}(E)\\ &\times G^{\sigma\; r}(E, \boldsymbol{k})^{\dagger}\Gamma^{\sigma}_\mathrm{L}(E, \boldsymbol{k}) G^{\sigma\; r}(E, \boldsymbol{k}) \Big\} \\
        & - \textnormal{Tr}\Big\{\big[f_{\textnormal{R}}(E)-F_{\textnormal{MB}}^{\sigma}(E,\boldsymbol{k})\big]\Gamma_{\textnormal{MB}}^{\sigma}(E)G^{\sigma\; r}(E, \boldsymbol{k})^{\dagger}\\
        & \times \Gamma^{\sigma}_\mathrm{R}(E, \boldsymbol{k}) G^{\sigma\; r}(E, \boldsymbol{k})\Big\}.
    \end{split}
\end{equation}
%
In DFT+NEGF, there is no many-body self-energy, and thus the incoherent contribution vanishes. The current 
is therefore solely given by Eq.~(\ref{eq: coherent current}), with $T^{\sigma}(E)$ evaluated using the KS Green's 
function, $g^{\sigma\; r}(E, \boldsymbol{k})$. Accordingly, charge transport is described as the coherent transmission 
of electrons through the central region and is determined by the occupation of the incoming and out-coming states, 
$f_\mathrm{L}$ and $f_\mathrm{R}$. The integrand in Eq.~(\ref{eq: coherent current}) is significantly different from 
zero only within an energy range approximately spanning from $E_\mathrm{F}-eV/2$ to $E_\mathrm{F}+eV/2$, 
referred to as the bias window. Thus, in practice, the current is often approximated by integrating $T(E)$ over this 
window.

Within DFT+NEGF+DMFT, both the coherent and incoherent contributions to the current may become 
significant. The coherent part, $I_{\textnormal{C}}$, still corresponds to the current arising from transmission through 
the central region, but is now evaluated using the many-body retarded Green's function, and therefore reflects a 
modified electronic structure compared to the DFT KS one. Importantly, due to the imaginary part of the many-body 
self-energy, the electronic states are not only shifted in energy but also acquire a finite lifetime. This already partially 
breaks the notion of perfectly coherent transport, despite the conventional terminology.
A similar effect has been reported within the Kubo formalism under the ``bare-bubble'' approximation implemented with 
DMFT Green's functions \cite{pi.ma.22}. In our calculations based on Eq.~(\ref{eq: coherent current}), this finite lifetime 
generally leads to a reduction of the transmission coefficient \cite{Ivan_Stefano_2009, andrea_Cu_co}, and consequently, 
of the current.

In contrast, the properly defined incoherent contribution to the current captures the impact of inelastic scattering 
on transport. Notably, Eq. (\ref{eq: incoherent current k dependent}) resembles Eq. (\ref{eq: coherent current}), 
but with the out-of-equilibrium electron distribution of the central region, $F_{\textnormal{MB}}^{\sigma}$, and the 
many-body broadening matrix, $\Gamma^{\sigma}_{\textnormal{MB}}$, replacing $f_\mathrm{R}$ ($f_\mathrm{L}$) 
and $\Gamma^{\sigma}_{\textnormal{R}}$ ($\Gamma^{\sigma}_{\textnormal{L}})$ in the first (second) term of the 
integrand. This suggests an intuitive picture in which the effect of electron-electron interactions can be represented 
by an additional ``virtual electrode'', characterized by the many-body self-energy and an effective distribution function $F_{\textnormal{MB}}(E, \boldsymbol{k})$. Then, $I_{\textnormal{IC}}$ can be interpreted as the current flowing 
from the left-hand side lead into this virtual electrode and subsequently from there into the right-hand side lead. 
Crucially, once electrons enter the virtual electrode, they may re-emerge with modified energy and phase, thereby 
mimicking inelastic scattering events arising from electron-electron interaction. This picture provides a natural analogy 
to the de-phasing B\"uttiker probe concept \cite{bu.85}, wherein phase-breaking and energy-dissipating processes 
are captured through a fictitious, current-conserving voltage probe \cite{ki.se.15}. In contrast, DMFT provides a 
quantitative, atomistic, and theoretically grounded framework to address this behavior.

The incoherent contribution to the current has often been neglected in literature. However, in some systems, 
it can have a significant impact, leading to a substantial increase in the total current with bias \cite{ne.sa.25}. 
In this work, we generally compute the total current using the original Meir-Wingreen formula and estimate 
$I_\mathrm{IC}$ by subtracting $I_\mathrm{C}$ from it, rather than applying 
Eq. (\ref{eq: incoherent current k dependent}), which is more complex to implement. 

\section{Computational Details}\label{section: computational details}
The DFT+NEGF calculations are performed using the LSDA \cite{ba.he.72,vo.wi.80} for the exchange-correlation 
functional. The leads' self-energies are calculated using the semi-analytic algorithm from Ref. \cite{ru.sa.08}. 
Norm-conserving Troullier-Martins pseudopotentials \cite{Tr.Ma.91} are employed to treat the core electrons. 
Valence states are expanded using a numerical atomic orbital basis set, which includes multiple-$\zeta$ and 
polarized functions \cite{so.ar.02}. The electronic temperature is set to 300~K, and the real space mesh is 
determined by an equivalent energy cutoff of 400~Ry.

At zero-bias, the DFT density matrix, given by Eq. (\ref{eq.DFT_rho}), is computed self consistently using a 
$20 \times 20$ transverse $\boldsymbol{k}$-point mesh. The integration over the energy is performed using 
a semi-circular contour in the complex energy plane \cite{ro.ga.06}. In this approach, 16 poles are used to 
represent the Fermi distribution, while 16 energy points are sampled along both the semi-circular arc and the 
imaginary line that forms the contour. The converged density matrix is read as an input for a non-self-consistent 
DFT calculation using a $200 \times 200$ transverse $\boldsymbol{k}$-mesh to obtain the zero-bias DFT DOS 
and transmission coefficient. This is enough for resolving the transmission function over the transverse BZ
in the case of tunneling. All energy values are shifted to align the Fermi level at 0~eV.

Finite-bias calculations are performed non-self-consistently, using the Hamiltonian of the central region 
calculated at zero-bias with an additional ramp potential \cite{Rudnev_Sci_Adv2017} that reproduces the 
linear voltage drop across the MgO insulating barrier (vacuum gap) in Fe/MgO/Fe MTJ (Cu/Co/vacuum/Co), namely $V_e (z) = -eV z/l + eV/2$ where 
$z=0$ ($z=l$) corresponds to the beginning (end) of the barrier (gap), as in Fig. \ref{fig:device}. Similar to the equilibrium 
case, a $200 \times 200$ transverse $\boldsymbol{k}$-mesh is used to obtain the finite bias DFT DOS and 
transmission coefficient.

DMFT calculations are performed assuming general orbital-dependent screened Coulomb interaction 
parameters for the $3d$ orbitals of Co and Fe. These parameters are expressed in terms of Slater integrals 
$F^0$, $F^2$ and $F^4$ (see reference [\onlinecite{im.fu.98}]). The ratio $F^4/F^2$ is assumed to correspond 
to the atomic value $\approx 0.625$ (Ref. [\onlinecite{an.gu.91}]). The average $U$ and $J$ interaction 
parameters are given through the relations $U=F^0$ and $J=(F^2+F^4)/14$.

The local Green's function in the DMFT loop \cite{andrea_Cu_co} is calculated by summing the retarded 
Green's function over $20 \times 20$ $\boldsymbol{k}$ points, whereas a $50 \times 50$ $\boldsymbol{k}$-mesh 
is considered to plot the DMFT DOS and transmission coefficient. The second-order contribution to the 
self-energy in the impurity solver is computed on an energy grid comprising $4000$ ($6000$) points and extending 
from $-10$~eV to $10$~eV ($-25$~eV to $15$~eV) at low bias for the Cu/Co/vacuum/Cu system (Fe/MgO/Fe MTJ).

\begin{figure}[t]
    \centering
      \includegraphics[clip,width=0.4\textwidth]{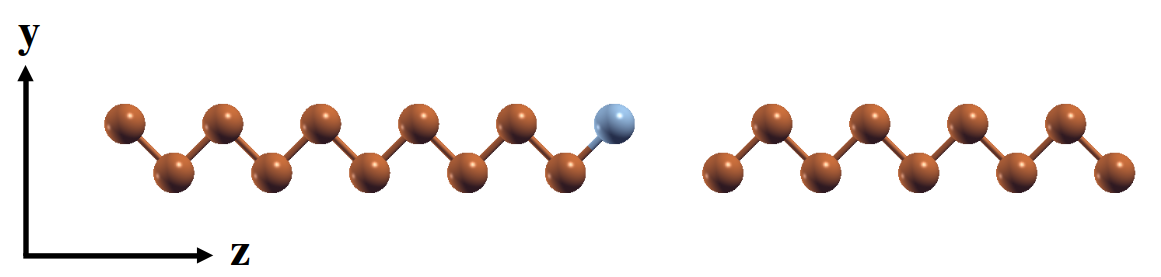}
   \caption{The Cu/Co device investigated in this work. A Co layer is attached to a left Cu lead and separated from 
   a right Cu lead by a 4 \AA~ vacuum gap. Periodic boundary conditions are applied in the plane transverse to the
   stack.}
    \label{fig: device+potential Cu/Co}
\end{figure}

\section{results} \label{section: results}
 \begin{figure*}[t!]
\centering
\includegraphics[width=0.9\textwidth]{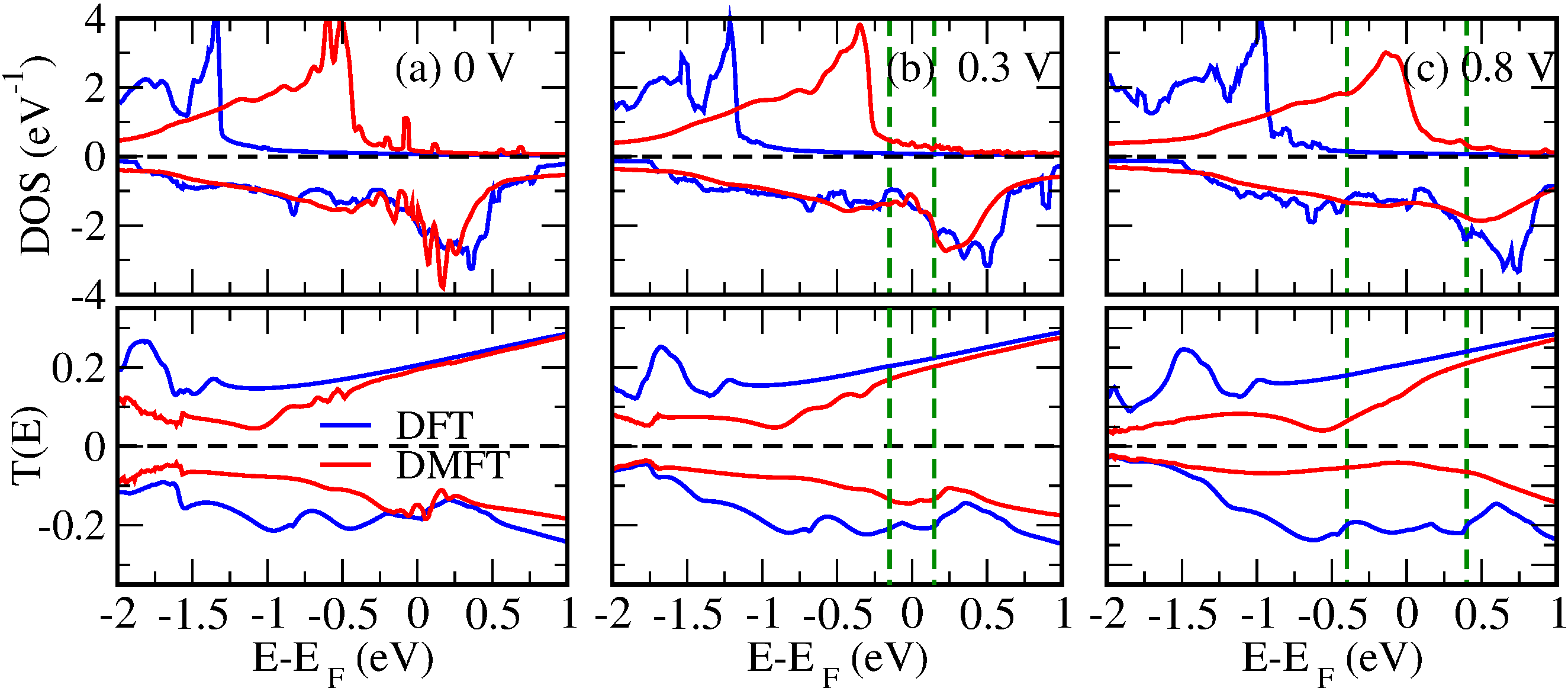} \caption{DFT and DMFT $3d$-Co PDOS and transmission coefficient 
at (a) $V=0$ V, (b) $V=0.3$ V, and (c) $V=0.8$ V. Spin-up (down) values are shown positive (negative). } \label{fig:trc} 
\end{figure*}

\begin{figure}[h]
\centering
\includegraphics[width=0.45\textwidth]{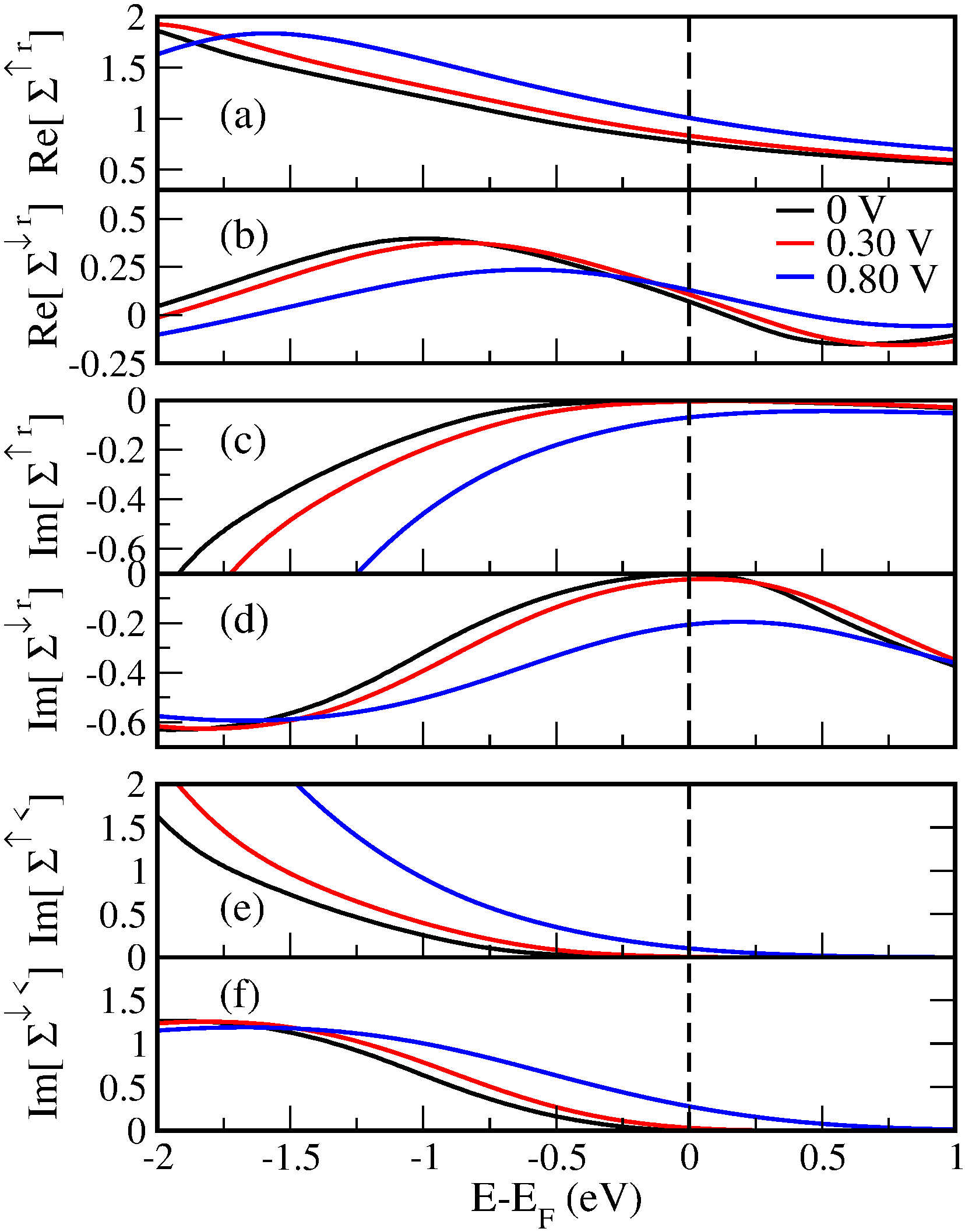} 
\caption{DMFT self-energy for Co, averaged over all the $3d$ orbitals, at $V= 0$ V, $V= 0.3$ V, and $V= 0.8$ V.
Top panel: real part of the retarded component. Middle panel: imaginary part of the retarded component.
Bottom panel: lesser component.} \label{fig: self_energy_bias} 
\end{figure}

\begin{figure}[h]
\centering
\includegraphics[width=0.45\textwidth]{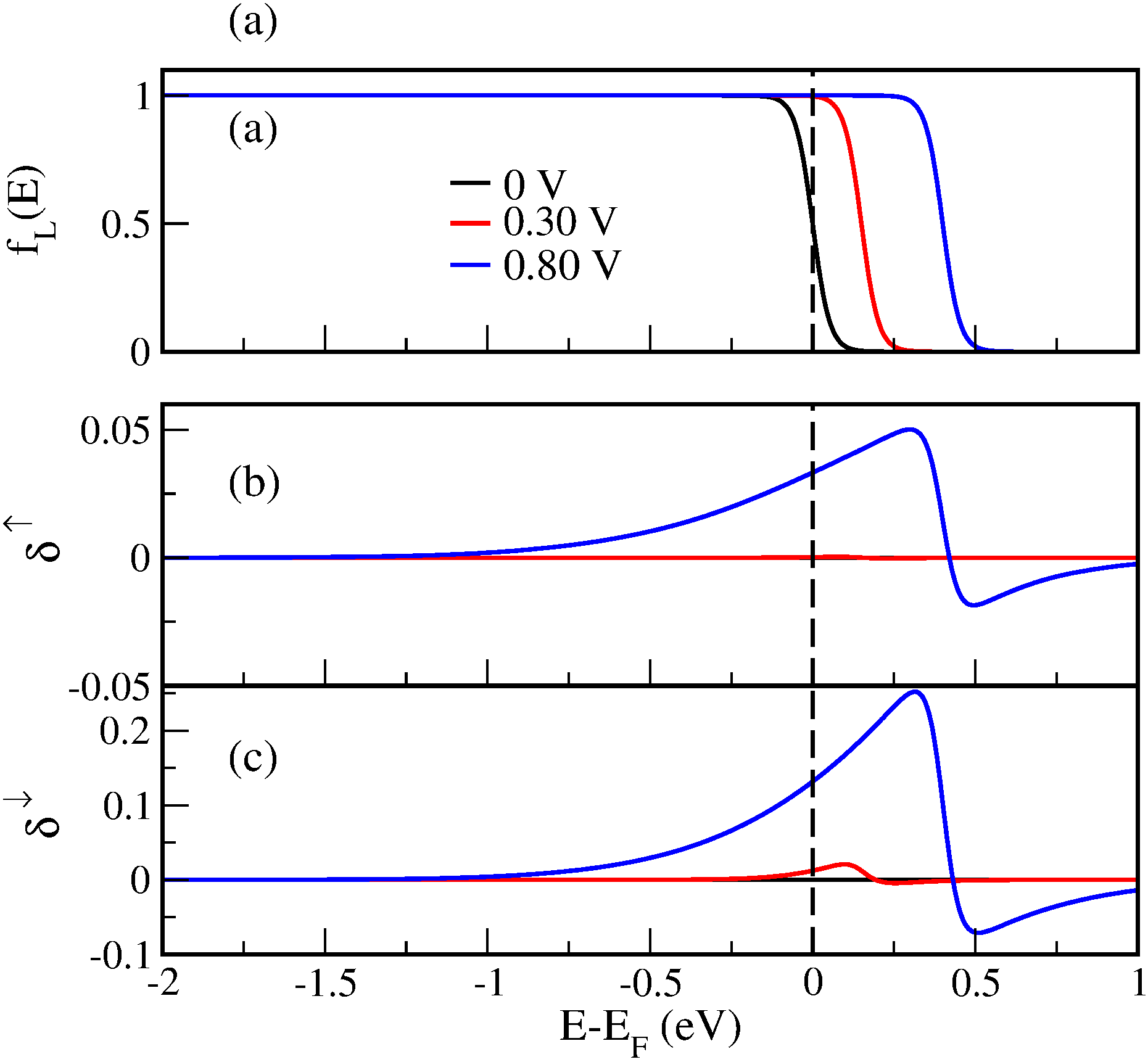} 
\caption{(a) Fermi function of the Cu left lead attached to the Co layer. In panel (b) and (c) 
we show the difference $\delta^\sigma = f(E, \mu_\mathrm{L}) - F^\sigma_\mathrm{MB}(E)$, 
averaged over all $3d$ orbitals, at $V = 0$ V, $0.3$ V, and $0.75$ V.
}
\label{fig: lesser_self_energy_bias_Cu_Co} 
\end{figure}

\subsection{Cu/Co/vacuum/Cu}\label{sec:CoCu}
The first system investigated comprises a Co mono-layer attached to a left-hand side Cu lead and 
separated from a right-hand side Cu lead by a 4~\AA-long vacuum gap, as shown in 
Fig. \ref{fig: device+potential Cu/Co}(a). At finite bias, the potential drops within this gap.
The setup mimics STM studies of magnetic surfaces. While a fully quantitative STM simulation 
would require the use of a large supercell with the right lead attached to a tip \cite{andrea_ivan_projection} 
--whose geometry and electronic structure may be nontrivial (e.g., \cite{Rudnev_Sci_Adv2017, dr.ru.20})-- 
the essential non-equilibrium physics related to the potential drop and charge current flowing through the 
central region is expected to be well captured already by our simplified system. 
The correlated subspace, $\mathcal{C}$ is spanned only by the Co $3d$ orbitals, whereas the 
Cu $3d$ orbitals are fully occupied and therefore play a negligible role over the correlation effects. 
We employ an average on-site Coulomb and exchange interaction parameters of $U = 1.5$ eV and 
$J = 0.5$ eV, respectively. 

\subsubsection{Zero-bias electronic structure}
The $3d$-Co PDOS calculated with DFT and DMFT at zero-bias are shown in Fig.~\ref{fig:trc}(a),
 illustrating the effect of static versus dynamical correlations. The DFT result (blue curve) exhibits a 
 pronounced spin-splitting: the spin-up states (majority states) are almost entirely occupied, with a 
 main peak at approximately $-1.3$ eV, whereas the spin-down states (minority states) extend from 
 low energies across $E_{\mathrm{F}}$ up to about $0.7$ eV. DMFT leads to a redistribution of the 
 PDOS (red curve), resulting in a reduced spin-splitting --- qualitatively similar to what has been reported 
 in many DMFT studies of transition metal ferromagnets \cite{br.mi.06, gr.ma.07, andrea_Cu_co, andrea_sigma_2, andrea_FeO, ne.sa.25}. This is the effect introduced by the real part of the retarded DMFT self-energy, shown in 
 Fig. \ref{fig: self_energy_bias} (averaged over all five $3d$ orbitals). In the spin-up channel, 
 $\text{Re}[\Sigma^{r\, \uparrow}_{\mathrm{avg}}(E)] $ is large and positive, pushing the PDOS upward 
 in energy toward $E_\mathrm{F}$. Conversely, in the spin-down channel, 
 $\text{Re}[\Sigma^{r\, \downarrow}_{\mathrm{avg}}(E)]$ becomes negative at $\sim 0.2$ eV, shifting the 
 high-energy part of the PDOS downward toward $E_\mathrm{F}$ and thereby sharpening its main peak.    

The imaginary part of the retarded DMFT self-energy vanishes at $E_\mathrm{F}$ and exhibits the typical 
behavior for a Fermi liquid, namely  $\textnormal{Im}[\Sigma_{\mathrm{avg}}^{r\,\sigma}(E)]\propto -(E-E_\mathrm{F})^2$ 
at low energies. Furthermore, it is related to the lesser DMFT self-energy via the fluctuation-dissipation relation 
in Eq. (\ref{eq: fd MB1}), reflecting the occupation of electronic states according to the Fermi-Dirac distribution 
in equilibrium.

\subsubsection{Finite-bias electronic structure}

Upon the application of a finite bias voltage $V$, the $3d$-Co PDOS shifts upward almost rigidly by 
$eV/2$, reflecting the strong coupling of the Co layer to the left lead. This effect is clearly visible in both 
the DFT and DMFT results shown in Figs. ~\ref{fig:trc}(b) for $V = 0.3$~V and \ref{fig:trc}(c) for $0.8$~V.
However, besides this somewhat trivial shift, the most interesting finding is that the imaginary part of the 
DMFT self-energy deviates from the Fermi-liquid behavior. Specifically, Im$[\Sigma^{r,\uparrow}_{\mathrm{avg}}(E_{\mathrm{F}})]$
becomes finite and increases with bias (Fig. \ref{fig: self_energy_bias}), indicating a loss of coherence due to 
electronic excitations. This displays in the progressive broadening and smoothing of the DMFT PDOS 
[Fig. \ref{fig:trc}(b) and (c)], particularly in the spin-down channel --- an effect that is, of course, absent in 
the effective single-particle DFT picture.

The departure from the Fermi-liquid regime at finite bias is also evident in the breakdown of the fluctuation-dissipation 
theorem. Specifically, the lesser DMFT self-energy $\Sigma^{<\;\sigma}_{\mathrm{avg}}(E)$, shown in 
Figs. \ref{fig: self_energy_bias}(e) and \ref{fig: self_energy_bias}(f), displays a strong deviation from Eq. (\ref{eq: fd MB1}) 
for energies within the bias window. This signals that the electronic distribution function departs from a simple 
Fermi function. To highlight this aspect, in Figs. \ref{fig: lesser_self_energy_bias_Cu_Co}(b) and 
\ref{fig: lesser_self_energy_bias_Cu_Co}(c) we plot $\delta^\sigma=f(E, \mu_\mathrm{L})-\bar{F}^\sigma_{\mathrm{avg},\mathrm{MB}}(E)$, 
where we take $f(E, \mu_\mathrm{L})$ [see Fig. \ref{fig: lesser_self_energy_bias_Cu_Co}(a)] as our reference 
equilibrium distribution, since the Co layer is much more strongly hybridized with the left-hand side lead, and $\bar{F}^\sigma_{\mathrm{avg},
\mathrm{MB}}(E)$ is the occupation of the correlated subspace averaged over the five $3d$ orbitals. 
The result shows that $\delta$ is positive for energies below $\mu_\mathrm{L}=eV/2$ where $f=1$, whereas 
it becomes negative for energies above $\mu_\mathrm{L}$, where $f=0$. This indicates that, away from equilibrium, 
the electronic distribution is further smeared out. In practice, electrons are excited from occupied states to higher-energy 
empty states, leaving behind holes. This effect becomes significant in particular for $V \gtrsim 0.4$, a threshold
that coincides with the entry of the spin-up DOS peak into the bias window, and directly reflects into the calculated 
current, as discussed in the following section.

\begin{figure}[b]
\centering
\includegraphics[width=0.45\textwidth]{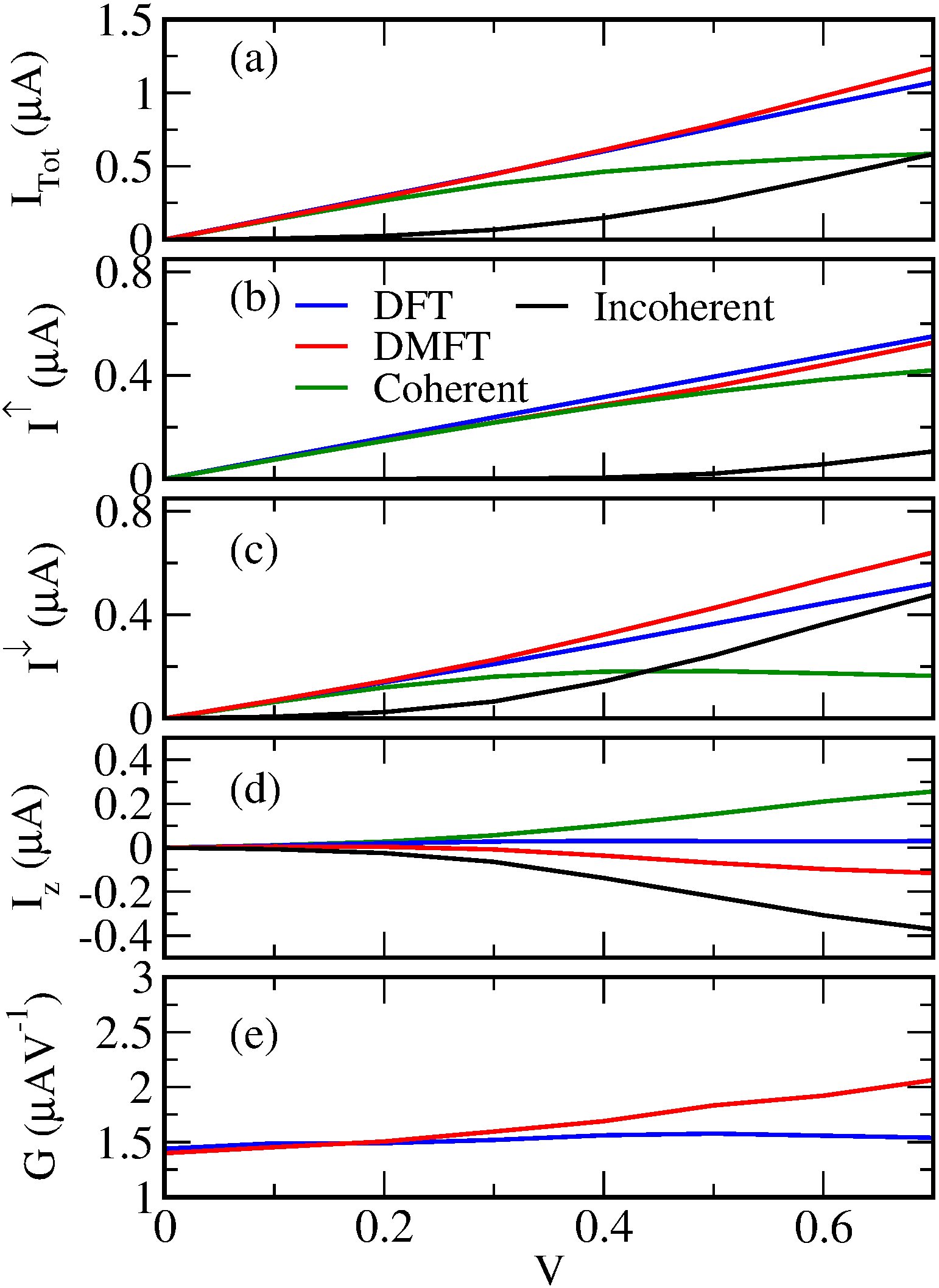} \caption{(a) Total, (b) spin up,  (c) spin down, (d) spin current and (e) conductance within  DFT and DMFT, decomposed into its coherent and incoherent contributions, as functions of the bias.} \label{fig: current} 
\end{figure}

\begin{figure}[b]
\centering
\includegraphics[width=0.5\textwidth]{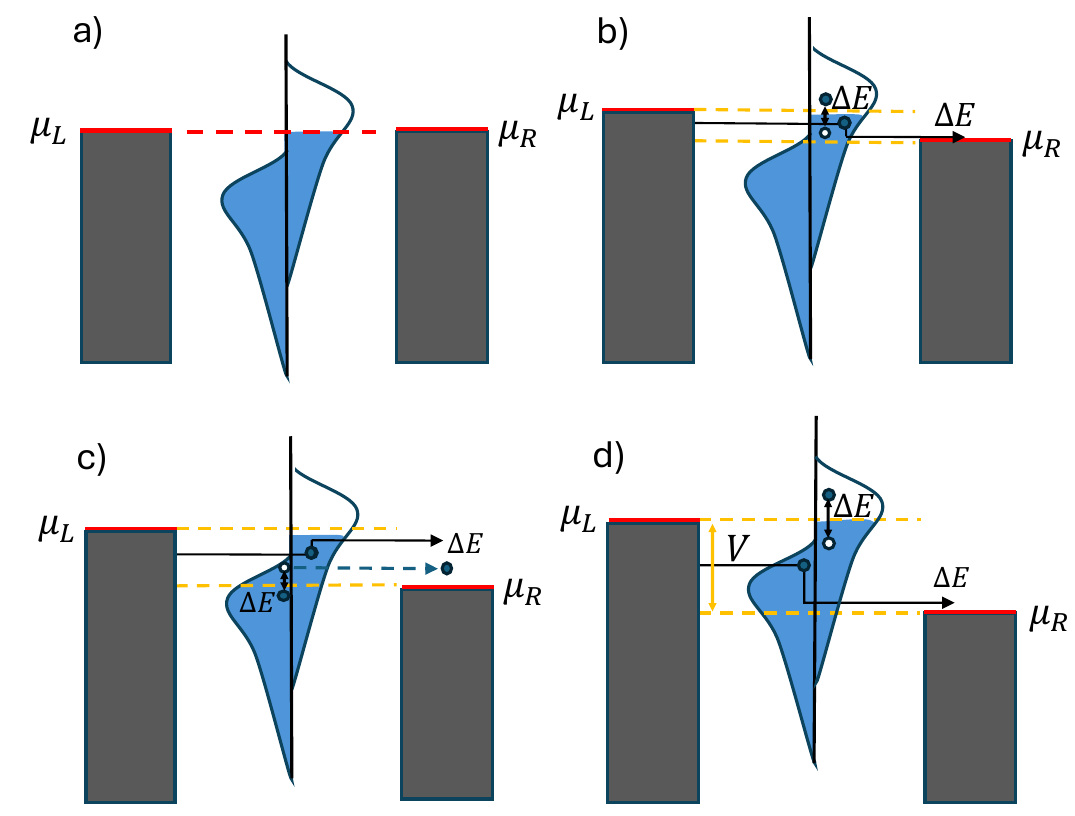} \caption{
Dominant scattering processes contributing to the incoherent current. Dark gray rectangles represent the Fermi boxes of the L and R leads, displaced by $\mu_\mathrm{L}-\mu_\mathrm{R}=V$, with the schematic Co $3d$ PDOS shown in between. The bias window is delimited by yellow dashed lines. Electrons (holes) are represented as filled (empty) circles.
(a) Zero-bias energy-level alignment. (b) A spin-down electron creates a spin-down electron–hole pair excitation and scatters into the right lead. (c) As the bias $V$ increases and the spin-up $3d$ states approach the bias window, virtual transfer of spin-up electrons into the right lead becomes possible (dashed blue arrow), leaving behind spin-up particle–hole pairs that can scatter with spin-down electrons. (d) Once the bias window is large enough to include a significant portion of the spin-up PDOS, spin-up electrons contribute directly to incoherent transport by exciting opposite- (or same-) spin electron–hole pairs and scattering inelastically into the right lead.} \label{fig: scattering} 
\end{figure}

\subsubsection{Current}
The charge current as a function of bias is plotted in Fig. \ref{fig: current}(a). The DFT and DMFT results 
(blue and red lines) coincide at low $V$ but begin to diverge for $V  \gtrsim 0.4$. Although the deviation 
remains relatively small, the underlying transport mechanisms differ fundamentally. In DFT the current is 
fully coherent, whereas in DMFT the incoherent contribution (black line) increasingly dominates with bias, particularly in the spin down channel.

The DFT current, obtained from the Landauer-B\"uttiker formula [Eq.~(\ref{eq: coherent current})] as the 
integral of the transmission coefficient over the bias window, increases linearly with $V$, since $T(E)$ is 
smooth and nearly constant across a wide energy range (see Fig.~\ref{fig:trc}). In the spin-up channel, 
transport is dominated by delocalized $4s$ states, which give a large transmission. In contrast, the $3d$-Co 
states lie well below the bias window for all $V$ considered, as seen in the PDOS in Figs. ~\ref{fig:trc}(b) 
and \ref{fig:trc}(c), and thus they are irrelevant for transport. In the spin-down channel, the $3d$-Co states 
play instead a more important role as they are located at energies near $E_\mathrm{F}$. However, their 
strong hybridization with the Cu left-hand side electrode causes a substantial spectral broadening, yielding 
a transmission coefficient that is equally large in magnitude and smooth in energy. Interestingly, this leads 
to spin-up and spin-down currents, which are nearly identical: namely, within the DFT picture, a single Co 
layer cannot spin-polarize the current in our set-up.

When the calculation is performed at the DMFT level, the coherent current (green line in Fig.~\ref{fig: current}) 
dominates at low bias and, as in DFT, increases linearly with $V$. At higher bias, however, it saturates, 
plateauing around 0.6~V, while the incoherent current rises sharply. This results in a crossover between 
coherent and incoherent transport at $V \sim 0.7$~V. Furthermore, the crossover is accompanied by the 
emergence of a negative spin current, $I^z = I^\uparrow - I^\downarrow$ [see Fig.~\ref{fig: current}(d)], 
indicating that bias-driven correlation effects induce spin-filtering.

The behavior of the DMFT coherent current arises from the reshaping of the transmission coefficient due 
to dynamical correlations [Fig.~\ref{fig:trc}], most notably the $\sim$0.8~eV upward shift of the spin-up $3d$ 
states, which brings them into the bias window already at $\sim$0.4~V. At the same time, the transmission 
through the Co $3d$ states is suppressed relative to the DFT result because of the damping and coherence 
loss of the electronic excitations (see 
the red curves in the bottom panels of Fig.~\ref{fig:trc}). This effect is encoded in the imaginary part of the DMFT self-energy. As $V$ increases and the system is driven further 
away from equilibrium and from a Fermi-liquid behavior, this suppression counterbalances the widening 
of the bias window, and eventually results in the coherent current plateau. The effect is stronger in the spin-down 
channel, where $3d$ states lie in the bias window even at low $V$, whereas the spin-up channel contributes 
only once the bias exceeds $\sim$0.4~V.

At a microscopic level, the suppression of coherent transport originates from bias-driven inelastic scattering 
processes, which conversely generate incoherent current. An electron entering the Co layer from the right-hand side lead
can scatter off another electron, creating an electron-hole pair excitation and transferring part of its energy, 
before being collected in an empty state of the right-hand side lead. The electron-hole pair excitations directly 
manifest in the non-equilibrium distribution function as already observed in Fig. \ref{fig: lesser_self_energy_bias_Cu_Co}. 
Scattering between electrons of opposite spin occurs with the full interaction strength $U$, whereas same-spin 
scattering is weaker, occurring with an effective interaction $U - J < U$, reduced because of the Hund's coupling. 
As a result, opposite-spin scattering processes dominate and produce the strongest effect on transport.

Several scattering processes can be specifically identified, as shown in Fig. \ref{fig: scattering}. A spin-down 
electron can create a spin-down electron-hole pair excitation [panel (b)]. Since the spin-down $3d$ states 
span the Fermi level, this process can occur even at low bias, resulting in a nonzero spin-down incoherent 
current already at small voltages, as seen in Fig.~\ref{fig: current}(c). As $V$ increases and the spin-up $3d$ 
states enter the bias window (approximately 0.35~eV), virtual transfer of spin-up electrons into the right lead 
becomes possible, leaving behind spin-up particle-hole pairs that can scatter with spin-down electrons 
[panel (c) in Fig. \ref{fig: scattering}]. Since this process involves opposite-spin interactions, it is stronger and 
leads to the sharper upturn in the spin-down incoherent current at $V\sim 0.3$~V seen in Fig.~\ref{fig: current}(c). 
Finally, at higher bias ($V\gtrsim 0.4$~V), once many spin-up states become available inside bias window, 
spin-up electrons likewise start contributing to incoherent transport by transferring energy to generate 
electron-hole pair excitations of both same and opposite spin [panel (d) in Fig. \ref{fig: scattering}]. This 
is reflected in the rise of the spin-up incoherent current in Fig. \ref{fig: current}(b). 

 Notably, the incoherent current contribution gives an increase in the differential conductance, $dI/dV$ at 
 $V \sim 0.25$ eV. This feature is consistent with STS measurements at positive bias on Co, though observed not 
in extended layers but at the centers of Co nanoislands on Cu \cite{PhysRevLett.99.246102, pa.pa.17}. While STS 
results are often interpreted in terms of coherent transport through the spin-up $3d$ Co bands within a DFT-based
Tersoff-Hamann picture \cite{PhysRevB.31.805}, our results indicate that this interpretation is incomplete. DMFT is 
crucial both to correctly position the spin-up $3d$ states, as DFT places them too low in energy, and to capture the 
relevant scattering processes.

While such processes have been included within the NEGF formalism for  $s–d$ exchange models \cite{hu.ba.11,PhysRevB.90.045115} and in {\it ab initio} approaches combining NEGF with time-dependent DFT for describing magnetic adatoms \cite{lo.co.10, sc.do.14}, they are usually neglected for surfaces. 
Our calculations demonstrate how these limitations can be overcome.

In summary, our analysis demonstrates that non-equilibrium correlations, captured by DMFT, qualitatively change 
the nature of the system, reshaping its electronic structure. The crossover between coherent and incoherent contributions 
under bias provides a natural explanation for experimental conductance features, calling for a revision of the commonly 
used single-particle pictures, and highlights the importance of many-body scattering in transport through magnetic layers.

\begin{figure*}[t]
\centering
\includegraphics[width=0.9\textwidth]{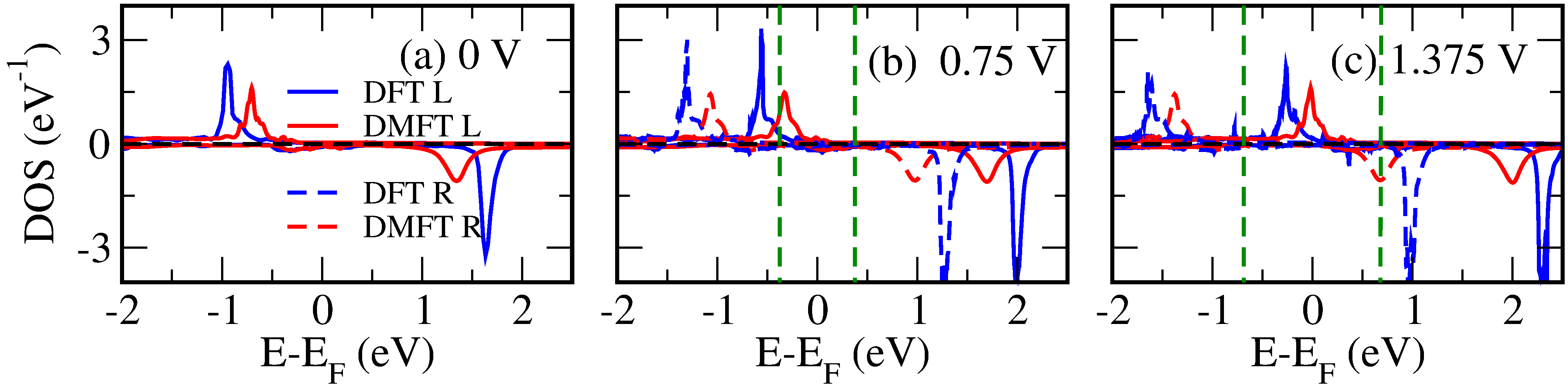} \caption{The DFT and DMFT PDOS of the $\Delta_{1}$ orbitals of the 
Fe layer to the left (L) and to the right (R) of the MgO barrier in Fe/MgO/Fe MTJ. Results are presented at (a) $V=0$~V, 
(b) $V= 0.75$~V and (c) $V=1.375$~V. Spin-up (down) values are shown positive (negative).
} \label{fig:DOS FeMgO} 
\end{figure*}

\subsection{Fe/MgO/Fe} \label{section: Fe/MgO}

The second system investigated is the prototypical Fe/MgO/Fe MTJ, which is modelled here by using the same set-up 
as in reference \cite{ne.sa.25}, shown in Fig. \ref{fig:projection}. The calculations presented in Ref. \cite{ne.sa.25} were based 
on the rigid-shift approximation of Sec.~\ref{sec:rigid_shift}, which was assumed valid {\it a priori}. In contrast, here we 
perform fully non-equilibrium calculations with our improved framework to explicitly test the validity of that approximation.

The CR of our MTJ consists of three Fe layers on each side of a six-layer MgO barrier and is connected to generic 
simple metallic leads, which are computationally simulated as a {\it bcc} lattice of Au atoms using only the $6s$ orbitals 
as a basis set. This effectively implements a broad-band approximation. The layers on the left-hand side of MgO 
can have a magnetization vector either parallel or antiparallel to that of the right-hand side layer, thereby realizing 
the two magnetic configurations of the MTJ. In this work, we focus exclusively on the parallel configuration and 
refer the reader to Ref. \cite{ne.sa.25} for a discussion of the antiparallel case and the TMR. The correlated subspace 
$\mathcal{C}$ comprises the Fe $3d$ orbitals, whereas the MgO barrier is treated within KS DFT, since it is weakly 
correlated. In fact, its conduction and valence bands have predominantly Mg-$3s$ and O-$2p$ character \cite{arya.92},
respectively. The DMFT calculations use the average screened interaction parameters $U =2.0$~eV and $J=0.5$~eV, 
which were found appropriate to describe Fe nanostructures in previous studies using our implementation of 
DFT+DMFT \cite{andrea_sigma_2}.

\subsubsection{Electronic structure at zero-bias}

Tunneling through Fe/MgO/Fe MTJs is dominated by spin-split evanescent states with $\Delta_1$ symmetry, 
which have the slowest decay rate inside the MgO barrier \cite{Bu_Zh+2001, Bu_2008}. Within our reference 
frame, where the transport direction is aligned along the $z$ axis, and for a cubic space group, the $\Delta_1$ 
states are those that transform as a linear combination of $s$, $p_{z}$, and $d_{z^2}$ atomic orbitals. 
The zero-bias $\Delta_1$-PDOS of the Fe atom at the Fe/MgO interface are presented in Fig.~\ref{fig:DOS FeMgO}(a) 
for DFT and DMFT, respectively. By symmetry, the results are identical for the two Fe layers at opposite side
of the junction.

Within DFT, sharp spin-split peaks appear at $E - E_\mathrm{F} \approx -0.9$~eV and $\approx 1.6$~eV for 
the spin-up and spin-down channels, respectively. In contrast, DMFT yields a substantial shift in energy of 
these peaks, resulting in a reduction of the spin-splitting, as observed in the Cu/Co system. This is due to 
the real part of the self-energy, and it is accompanied by a pronounced broadening caused by the imaginary 
part, which vanishes at $E_\mathrm{F}$ as a consequence of its Fermi-liquid nature. Our results reproduce 
exactly those reported in Ref.~\cite{ne.sa.25}, except for small quantitative differences in the peak positions 
that can be attributed to the different treatment of the double-counting correction [see discussion in 
Sec. \ref{section: impurity solver}]. Overall, our results provide a detailed account of correlation effects at the 
Fe/MgO interface.

 \begin{figure}[t]
\centering
\includegraphics[width=0.45\textwidth]{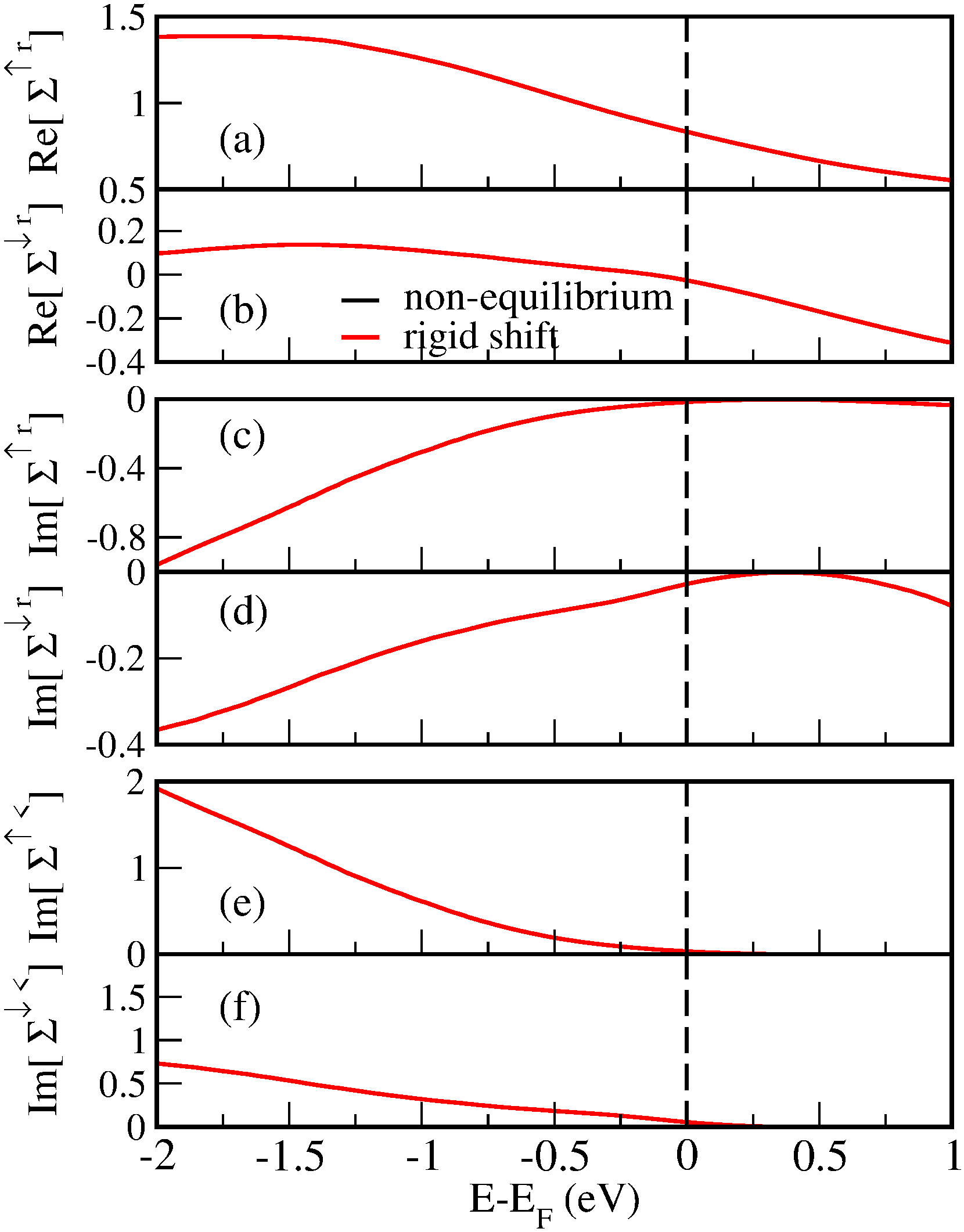} 
\caption{Comparison of the DMFT self-energy for the Fe monolayer closest to the left-hand side of the MgO barrier. 
Results are averaged over all $3d$ orbitals, in the Fe/MgO/Fe MTJ with and without the rigid-shift approximation 
at $V = 0.75$~V. Top panel: real part of the retarded component. Middle panel: imaginary part of the retarded component.
Bottom panel: lesser component.}
\label{fig:lesser_self_energy_Fe_MgO} 
\end{figure}

\subsubsection{Electronic structure at finite-bias}

At  finite bias, the electrostatic potential drops within the MgO barrier. The Fe $\Delta_1^\sigma$ state at the 
left-hand side Fe/MgO interface, $\Delta_{1(\mathrm{L})}^\sigma$, shifts upward in energy by $+eV/2$, while the corresponding 
state at the right-hand side, $\Delta_{1(\mathrm{R})}^\sigma$, shifts downward by $-eV/2$, as  can be seen in Fig.~\ref{fig:DOS FeMgO}(b) 
and \ref{fig:DOS FeMgO}(c). Differently from the Co/Cu case, where the self-energy undergoes significant changes at 
large bias, in the Fe/MgO/Fe MTJ the imaginary part of the self-energy [Fig.~\ref{fig:lesser_self_energy_Fe_MgO}(c) and 
\ref{fig:lesser_self_energy_Fe_MgO}(d)] just shifts with energy, essentially retaining its Fermi-liquid character, to vanish 
quadratically at $\mu_\mathrm{L}=E_\mathrm{F}+eV/2$ ($\mu_\mathrm{R}=E_\mathrm{F}-eV/2$) for the Fe atoms at the left-hand (right-hand) side interface.

Since the correlated Fe atoms are strongly coupled to one lead and separated from the opposite lead by a sufficiently 
thick MgO barrier --- so that their interaction with the latter is negligible --- they can be considered in equilibrium with 
the lead which they are coupled to. Indeed, we find that the retarded and lesser self-energies satisfy the fluctuation-dissipation 
relation, Eq.~(\ref{eq: fd MB1}), at all $V$, as shown in Fig. \ref{fig:lesser_self_energy_Fe_MgO}. This directly demonstrates 
that the rigid-shift approximation holds for this system and confirms {\it a posteriori} its application in Ref. \cite{ne.sa.25}. 
As a result, the electronic structure at finite bias essentially follows from the zero-bias one, with genuine non-equilibrium 
correlation effects being negligible.

\subsubsection{Current}

\begin{figure}[h]
\centering
\includegraphics[width=0.45\textwidth]{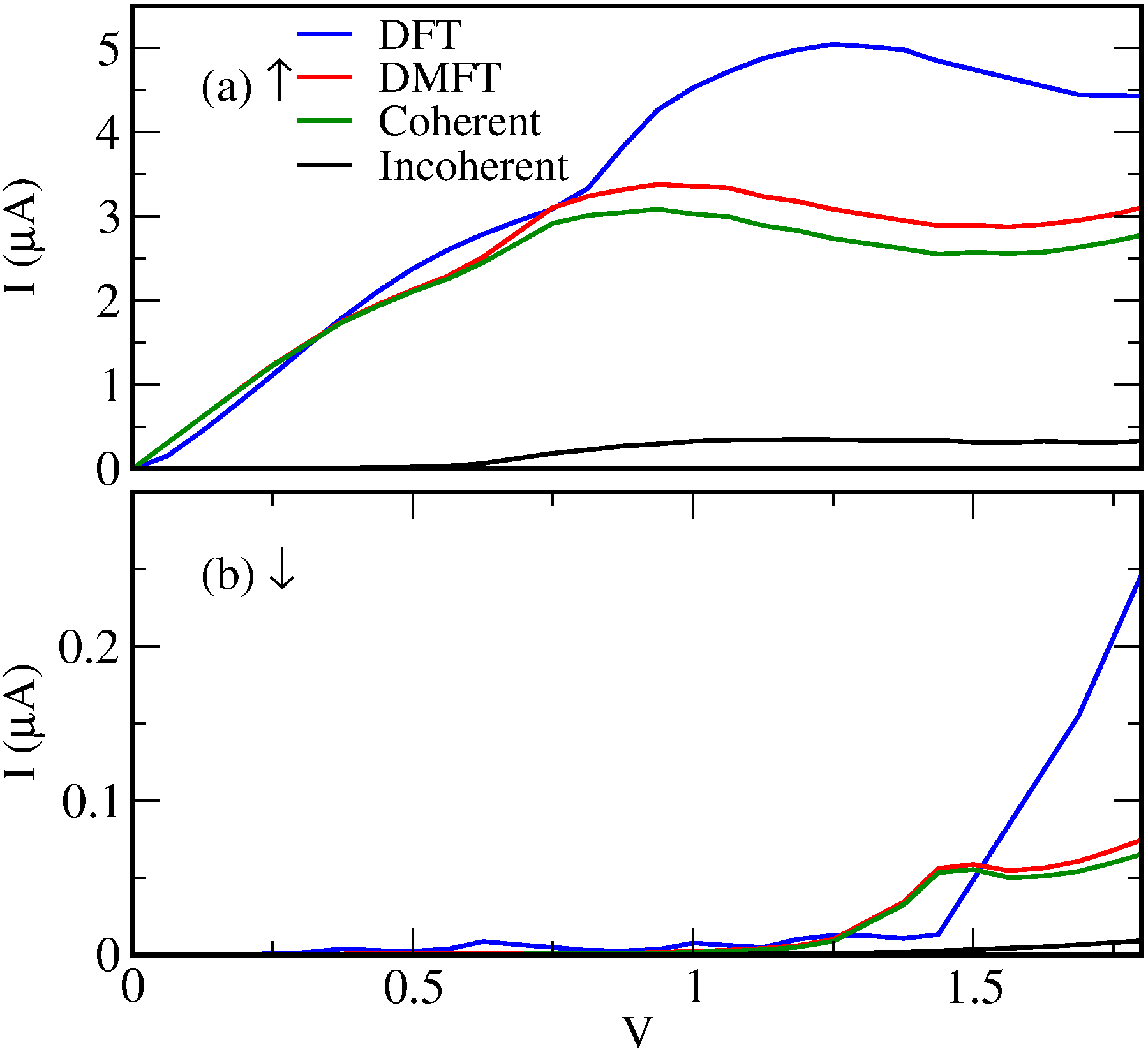} \caption{Spin (a) up and (b) down currents for the Fe/MgO/Fe MTJ 
calculated with DFT and DMFT, and decomposed into coherent and incoherent contributions as a function of bias.} 
\label{fig: charge and spin current FeMgO} 
\end{figure}

The charge current as a function of the bias is shown in Fig. \ref{fig: charge and spin current FeMgO}.
At low biases, the DFT and DMFT results nearly coincide, with a fully coherent current dominated by the spin-up 
component, $I^\uparrow$. This is because transport is mostly through the highly conductive $s$ states forming 
the Fe $\Delta_1$ band. As the bias increases, however, the spin-up $\Delta_1$ peak --- dominated by $d_{z^2}$ 
states from the left Fe layers --- enters the bias window [Fig.~\ref{fig:DOS FeMgO}(b) and \ref{fig:DOS FeMgO}(c)], 
and $I^\uparrow$ saturates into a plateau. The crossover from the linear regime to the plateau occurs in DMFT at 
a $V$ lower than in DFT, reflecting the different energy positions of the spin-up $d_{z^2}$ states predicted by the 
two methods.
More interestingly, at this crossover bias, DMFT also reveals the onset of an incoherent contribution to the current, 
similar to what we found in the Cu/Co junction. In this case, however, the incoherent component accounts for only 
about 10\% of the total spin-up current. The underlying scattering process involves an incoming spin-up electron 
interacting with electron-hole pair of the same spin, which is activated when the spin-up $\Delta_1$ state of the left 
layers enters the bias window, at $V \approx 0.7$ V.

In the spin-down channel, the current remains negligible as long as the spin-down $\Delta_1$ state is outside the 
bias window. Once this state enters the window, the spin-down current suddenly increases, with the threshold 
in DMFT being at lower energies than in DFT. This, again, reflects the DMFT shift in the energy of the $\Delta_1$ state.

In summary, although the applied bias has only a weak effect on Fe, bias-induced renormalization of the 
$\Delta_1$ energy position and scattering from dynamical correlations remain important in Fe/MgO/Fe. This 
is consistent with the findings of Ref. \cite{ne.sa.25}, where their experimental relevance is discussed in detail. 
DMFT thus offers a better description than DFT, even for this technologically relevant system.

\section{Conclusion} \label{section: conclusion}

We have introduced a framework that combines DFT, DMFT and the NEGF method to study steady-state 
transport in two-terminal devices beyond the linear response, and including dynamical electron-correlation effects. 
By employing second-order perturbation theory as impurity solver, our approach is well suited for moderately 
correlated $3d$ transition metals. However, the method can in principle be extended to more strongly correlated 
materials, provided that reliable multi-orbital non-equilibrium solvers become available --- an outstanding challenge 
at present.

As a first application, we have investigated a single Co monolayer sandwiched between Cu electrodes, where 
a vacuum gap is introduced at one side to enable a potential drop under bias. This is a configuration that resembles
that encountered in an STM/STS experiment. Our results show that non-equilibrium dynamical correlation in Co drives 
a qualitative transition from Fermi-liquid to non-Fermi-liquid behavior, originating from inelastic scattering of electrons 
with collective electron-hole excitations. This re-shapes the electronic structure and generates incoherent contributions 
to the current, accompanied by a sharp increase in conductance around $V \sim 0.4$~V, where the spin-up Co states 
enter the bias window. These findings provide a natural explanation for STS results, calling for a revision of commonly 
used single-particle models, and underscore the critical role of many-body scattering in transport through magnetic layers.

We have then applied the framework to an Fe/MgO/Fe MTJ. We find that the Fe layers remain close to equilibrium, 
showing that the rigid-shift approximation of Ref. \cite{ne.sa.25} holds for this system. As a result, the electronic structure 
at finite bias essentially follows that at zero bias, with genuine non-equilibrium correlation effects being negligible. Nonetheless, 
we still observe the emergence of a partially incoherent current as the Fe $3d$ states enter the bias window, in agreement 
with previous results \cite{ne.sa.25}, demonstrating that dynamical correlation affects transport even in this technologically 
relevant system.

Overall, our framework provides a pathway toward modeling spintronic devices with explicit many-electron effects 
at finite bias in far-from-equilibrium conditions. Looking ahead, this can be generalized to compute spin currents in 
non-collinear and spin-orbit-coupled systems by building on the computational modules already employed in standard 
DFT+NEGF calculations \cite{book1,ba.gu.24}. This will make it possible to study how scattering processes, such as 
those identified here, influence spin-transfer torque \cite{Slonczewski96, Berger96} and spin-orbit torque \cite{Miron2011,Li.Pa.12}, 
offering new insights into current-induced spin dynamics beyond the state-of-the-art {\it ab initio} single-particle 
treatment.

\section*{Acknowledgements}
D.N. was supported by the Irish Research Council (Grant No. GOIPG/2021/1468 ). A.D. acknowledges funding 
by Science Foundation Ireland (SFI) and the Royal Society through the University Research Fellowship URF/R1/191769 
during the initial stage of the project, when he was employed at Trinity College Dublin. 
S.S. thanks Research Ireland (AMBER Center grant 12/RC/2278$_-$P2) for support.
Computational resources were provided by Trinity College Dublin Research IT.

\appendix
\renewcommand{\thesection}{A\arabic{section}} 
\renewcommand{\thefigure}{A\arabic{figure}}  
\setcounter{figure}{0}

\section{Current conservation} \label{section: current conservation}

\begin{figure}[h]
\centering
\includegraphics[width=0.45\textwidth]{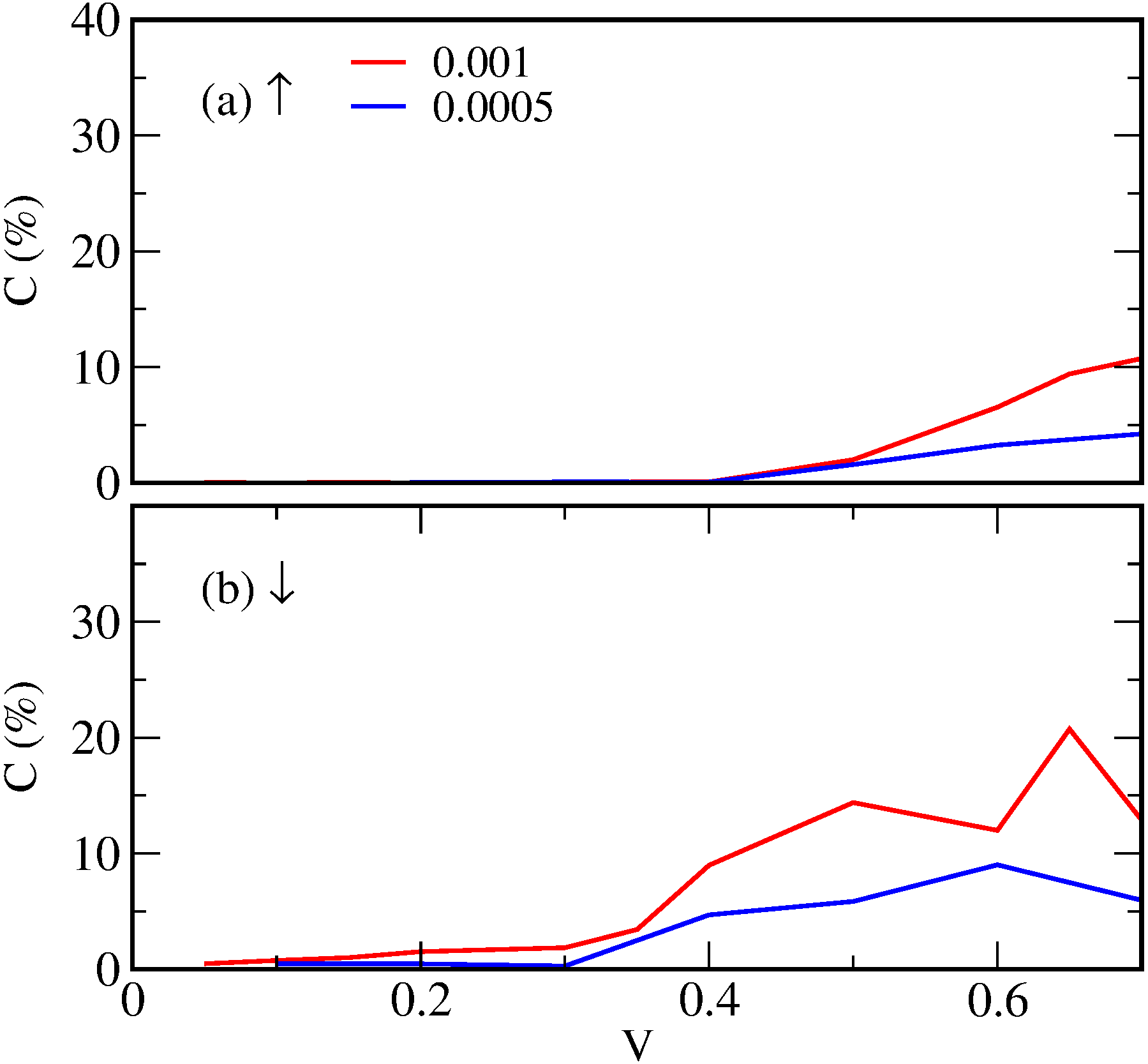} \caption{The current conservation for the (a) spin up and (b) spin down channel.  } \label{fig: current conservation} 
\end{figure}

Charge conservation in DFT+NEGF+DMFT is quantified as $C = 100\%  (2 |\frac{I_\textnormal{L}-I_{\textnormal{R}}}{I_{\textnormal{L}}+I_{\textnormal{R}}}|)$. Figure~\ref{fig: current conservation}  shows $C$ for DMFT self-energies converged in the self-consistent loop to tolerances of $10^{-3}$ and $5\times10^{-4}$. Reducing the tolerance improves current conservation across the bias range. The data in Fig.~\ref{fig: current} use the $5\times10^{-4}$ tolerance, yielding a maximum deviation of ~10\%. This demonstrates that our DFT+NEGF+DMFT framework provides qualitatively --- and potentially quantitatively --- reliable predictions for transport in realistic heterostructures.

\bibliography{FeMgO}

\end{document}